\documentclass[preprint,12pt]{aastex}
\newcommand{\torb}{t_{\rm orb}}

\slugcomment{$$Id: ms.tex 632 2007-07-11 20:55:22Z joishi $$}
\begin{document}

\title{Turbulent Torques on Protoplanets in a Dead Zone}
\author{Jeffrey S. Oishi\altaffilmark{1,2}, 
  Mordecai-Mark Mac Low\altaffilmark{2}, Kristen Menou\altaffilmark{3}}
\altaffiltext{1}{Department of Astronomy, University of Virginia,
  P.O. Box 3818, Charlottesville, VA 22903; Email: joishi@amnh.org} 
\altaffiltext{2}{Department of Astrophysics, American Museum of
  Natural History, Central Park West at 81st St, New York, NY
  10024-5192; Email: mordecai@amnh.org} 
\altaffiltext{3}{Department of Astronomy, Columbia University, 550
  West 120th St, New York, NY 10027; Email: kristen@astro.columbia.edu}

\begin{abstract}
Migration of protoplanets in their gaseous host disks may be largely
responsible for the observed orbital distribution of extrasolar
planets. Recent simulations have shown that the magnetorotational
turbulence thought to drive accretion in protoplanetary disks can
affect migration by turning it into an orbital random walk. However,
these simulations neglected the disk's ionization structure. Low
ionization fraction near the midplane of the disk can decouple the
magnetic field from the gas, forming a dead zone with reduced or no
turbulence. Here, to understand the effect of dead zones on
protoplanetary migration, we perform numerical simulations of a small
region of a stratified disk with magnetorotational turbulence confined
to thin active layers above and below the midplane. Turbulence in the
active layers exerts decreased, but still measurable, gravitational
torques on a protoplanet located at the disk midplane. We find a
decrease of two orders of magnitude in the diffusion coefficient for
dead zones with dead-to-active surface density ratios approaching
realistic values in protoplanetary disks. This torque arises primarily
from density fluctuations within a distance of one scale height of the
protoplanet.  Turbulent torques have correlation times of only $\sim
0.3$ orbital periods and apparently time-stationary
distributions. These properties are encouraging signs that stochastic
methods can be used to determine the orbital evolution of populations
of protoplanets under turbulent migration. Our results indicate that
dead zones may be dynamically distinct regions for protoplanetary
migration.
\end{abstract}

\keywords{accretion --- MHD --- planetary systems: formation ---
planetary systems: protoplanetary disks --- turbulence}

\section{Introduction}

Detections of extrasolar planets now number in the hundreds
\citep{Betal06}. Their orbital parameters reveal a number of
multi-planet systems in mean motion resonances
\citep{LP02,JKLL03,LBFMV06} and many Jupiter and Neptune-mass objects
with major axes under an astronomical unit \citep{GDGU06}. These
observations can be explained by orbital migration arising from
disk-planet interaction \citep{LP79,GT80}.

Protoplanetary migration driven by gravitational interaction with a
disk can occur in at least three ways. Type I migration occurs when
protoplanets with masses $\lesssim 10-30 {\rm M}_\oplus$ raise density
waves in the disk, which in turn exert torque on the protoplanet
\citep{GT80}. This migration is usually inward due to the radial
gradients in pressure, temperature, and sound speed from the global
structure of typical disks \citep{W86}. Type II migration is an
essentially non-linear process that occurs when a massive planet opens
a gap in the surrounding accretion disk \citep{W97}. The planet then
couples to the viscous evolution of the disk, migrating inward on the
local disk accretion timescale. Type III migration occurs when
material in the coorbital region of a roughly Saturnian mass object
begins to exert strong corotation torque that scales with the
migration rate, leading to runaway \citep[][though the existence of
runaway is questioned by \citet{DBL05} on resolution
grounds]{MP03}. We will here be solely concerned with Type I
migration.

For an $\sim 1 \mathrm{M}_\oplus$ object at $5\mbox{ AU}$, the time
scale for Type I migration is $\tau_{mig} \sim 8 \times 10^5 \mbox{
yr}$ \citep*[e.g.,][]{TTW02}, while protoplanetary disks have
lifetimes constrained to be less than $t_{disk} \sim 10^7 \mbox{ yr}$
\citep[e.g.,][]{JCSBv06}. This poses a significant challenge to the
core-accretion scenario for gas giant planets \citep{PHBLPG96}, in
which rocky cores take $\sim 10^7 \mbox{ yr}$ to grow to $\sim 10-30
\mathrm{M}_\oplus$ before rapidly accreting a gaseous
envelope. \citet{RA03} note that if a core migrates stochastically,
the timescale for core accretion drops by an order of magnitude from
this estimate. More recent studies \citep*{HBL05} show a core-accretion
timescale of $1-5 \times 10^6 \mbox{ yr}$ for Jupiter, still
considerably longer than the migration timescale.

Essential to this discussion of protoplanetary migration is the disk
itself. Disks around young stars are known to be accreting
\citep{HCGD98}.  Angular momentum must be transported outward to allow
accretion. The strength of the transport mechanism can be most simply
parameterized by the dimensionless viscosity $\alpha = \nu_t /(c_s H)$
\citep{SS73}.  Models incorporating a viscosity characterized by
$\alpha$ are known as $\alpha$-disks.  Numerical simulations of
disk-planet interaction confirm the basic analytic predictions of the
timescales for Type I migration in purely smooth, featureless, laminar
disk models including $\alpha$-disks and the minimum mass solar nebula
\citep*{MTTI99,M02,DKH03}.

However, protoplanetary accretion disks are far from featureless,
laminar disks. Recently, two separate approaches to more complex
modeling of planet-disk interaction have been identified and
pursued. First, more sophisticated one-dimensional $\alpha$-disk
models with radial opacity jumps \citep{MG04} and departures from
isothermality \citep{JS05} show that Type I migration rates can be
significantly altered by regions of long migration time that function
as traps. Additionally, \citet{PM06} have performed three-dimensional
radiation hydrodynamics simulations of Type I migration and found that
Type I migration can be halted and even reversed if the disk cannot
radiate efficiently in the corotation region of the planet's
orbit. Second, angular momentum transport most likely is actually
driven by some form of turbulence, likely the saturated state of the
magnetorotational instability (MRI) \citep[see][for a
review]{BH98}. Understanding the effects of turbulence on migration
requires explicit, three-dimensional, magnetohydrodynamic (MHD)
simulations. In a series of recent papers, Nelson, Papaloizou, and
collaborators have performed such simulations including the effects of
a protoplanet embedded within the turbulent flow (\citealt*{PN03,
NP03, PNS04, NP04, N05}, see also \citealt*{LSA04}). These models
revealed that for an object with mass $\lesssim 30 \mathrm{M}_\oplus$,
torque fluctuations from turbulent density perturbations cause the
protoplanet to follow a random walk, possibly significantly
lengthening its lifetime.

\citet*[hereafter JGM06]{JGM06} have argued, using semi-analytic
models, that these fluctuations will \emph{decrease} the lifetimes of
most planets, but allow a small number of them to scatter to large
radii and thus survive. Their approach is based on a Fokker-Planck
formalism that treats Type I migration as advection and the orbital
motions induced by stochastic torques from turbulence as an
independent diffusion process. Such a model depends on details of the
turbulence that can only be provided by numerical simulation.

Because protoplanetary disks are cold and dense at their midplanes,
the ionization fraction is a strong function of height.  The MRI is
stabilized by Ohmic diffusion for magnetic Reynolds numbers $Re_M
\lesssim 10^4$ for zero-net flux poloidal fields, though the actual
value depends strongly on the geometry of the field \citep*{FSH00} and
possibly the presence of artificial resistivity
\citep{BDH04}. Therefore, the MRI can not operate at all heights at
radii where planet formation likely occurs \citep{G96}. Modeling of
the ionization structure of the disk leads to a three layer
description \citep{G96,SMUN00,FTB02,SWH04,IN06} with MRI turbulence
occurring in active zones above and below a dead zone at the midplane.

Our purpose in this paper is twofold. First, we wish to understand the
effect of a dead zone on the migration of a low-mass protoplanet at
the midplane. Second, we wish to further quantify the torques on a
protoplanet caused by MRI turbulence, with and without dead zones, in
order to verify the assumptions of and provide better parameters for
the Fokker-Planck model of JGM06.

\S~\ref{sec:models} describes our models, \S~\ref{sec:plasma} gives a
brief overview of the turbulence and dead zones in the plasma,
\S~\ref{sec:torques} provides analysis of the torques on protoplanets
in real and Fourier space, \S~\ref{sec:discussion} discusses the
applicability of our results to real disks and statistical treatments
of ensembles of migrating protoplanets, and \S~\ref{sec:conclusions}
presents our conclusions.

\section{Models} \label{sec:models}
Torques produced by turbulent density perturbations compete with Type
I migration in the turbulent migration scenario proposed by
\citet{N05}.  In order to focus on the characteristics of the
turbulence, we exclude active Type I migration in our models by
considering a zero-mass test particle fixed at the center of the
box. A test particle does not raise density waves in the disk,
allowing us to isolate the torques on the planet from disk
turbulence. We work in a local frame, allowing for maximum resolution
of the disk near the planet---where, as we will show, most of the
turbulent torque originates. Our test particle approach has the
additional advantage of allowing us to study turbulent migration in
the local frame without worrying about spurious density wakes from the
(shearing) periodic boundary conditions we employ
\citep[eg,][]{PNS04,NP04}. However, in doing so, we implicitly assume
that Type I and turbulent torques are separable and additive. It is
not clear that this assumption is valid, as \citet{N05} found that
explicit Type I migration in turbulent models differed from the linear
combination of stochastic migration on massless protoplanets and a
Type I migration rate from laminar disk calculations, suggesting that
Type I migration is itself modified by MRI turbulence.

\subsection{Numerical Method}
We use the Pencil Code\footnote{available at
http://www.nordita.dk/software/pencil-code/}, a spatially sixth-order
and temporally third-order finite difference MHD code \citep{BD02}
designed to study weakly supersonic turbulent flows. The Pencil Code
solves the MHD equations in non-conservative form, monitoring
conservation as a check on the quality of the solutions. 

We work in the shearing box approximation \citep*[e.g.,][]{GL65,
HGB95}, in which one considers a small Cartesian box around a fiducial
point in the disk at radius $R$, and expands the gravitational
potential from the central protostar to leading order in $H/r$. The
coordinate frame is oriented such that $x$ is the local radial
direction, $y$ points in the azimuthal direction, and $z$ is mutually
orthogonal. We solve for the departures $\mathbf{u}$ from the mean
Keplerian shear flow, $u_y^{(0)} = -3/2 \Omega x$. The shearing box
MHD equations are thus
\begin{equation}
\label{continuity}
\partial_t \rho + u_y^{(0)} \partial_y \rho + \mathbf{\nabla} \cdot (\rho \mathbf{u}) = f_{\rho}
\end{equation}
\begin{equation}
\label{eqmotion}
\partial_t \mathbf{u} + \mathbf{u} \cdot \mathbf{\nabla}
\mathbf{u} + u_y^{(0)} \partial_y \mathbf{u} = -\frac{\mathbf{\nabla}P}{\rho} -
 \mathbf{\nabla} \Phi+ \frac{\mathbf{J \times B}}{\rho}
 + \mathbf{f_u} + 2 \Omega u_y \mathbf{\hat{x}} - \frac{\Omega
   u_x}{2}\mathbf{\hat{y}} - \Omega^2 z \mathbf{\hat{z}} + \zeta (\nabla \nabla\cdot \mathbf{u})
\end{equation}
\begin{equation}
\label{induction}
\partial_t \mathbf{A} + u_y^{(0)} \partial_y \mathbf{A} = \frac{3
  \Omega A_y}{2}\mathbf{\hat{x}} + \mathbf{u \times B} + \eta(z) \mu_0 \mathbf{J} + \mathbf{f_A}
\end{equation}
where $\Omega$ is the rotation rate of the box,
$f_{\rho}$, $\mathbf{f_u}$, $\mathbf{f_A}$ are stabilizing
hyperdiffusivities (see below), $\mathbf{A}$ is the magnetic vector
potential, $\mathbf{J = \nabla \times B}/\mu_0$ is the current
density, $\eta(z)$ is a height-dependent resistivity, and the other
symbols have their standard meaning. The extra term $u_y^{(0)}
\partial_y$ in all dynamical equations is a result of
subtracting the Keplerian shear from the velocity, as is the $(3/2)
\Omega A_y$ term in equation~(\ref{induction}). Finally, we close the
set with an isothermal equation of state,
\begin{equation}
\label{eqstate}
P = c_s^2 \rho,
\end{equation}
where $c_s$ is the isothermal sound speed.

Because the Pencil Code uses spatially centered finite differences,
there is no formal diffusion in the algorithm, so explicit diffusion
operators must be included in all dynamical equations in order to
ensure numerical stability. In this work, we use sixth order
hyperdiffusion for continuity, momentum, and induction, in addition to
physical Laplacian (i.e., second order) diffusion. Sixth order
hyperdiffusion has the form $\nu_6 \mathbf{\nabla^6 u}$ and damps
signals only near the Nyquist wavenumber at the smallest scales of the
box, allowing larger modes to remain largely unaffected. Our
formulation conserves mass and momentum by construction. For details,
we refer the reader to \citet{JK05}.  We capture shocks using a von
Neumann artificial viscosity implemented as a bulk diffusion with
variable coefficient $\zeta = c \langle \mathrm{max} [\mathbf{( -
\nabla \cdot u})_+] \rangle $ (where the max is taken over $3$ zones)
on the continuity and momentum equations \citep*{HBM04}.

For our dead zone runs, we have modified the Pencil Code to include a
height-dependent resistivity with the profile given by \citet{FS03}:
\begin{equation}
\label{resistivity}
\eta(z) = \eta_0 \exp\left(-\frac{z^2}{2}\ \right)
\exp\left(\frac{\Sigma_0}{\Sigma_{CR}} \frac{1}{2\sqrt{\pi}} \int_z^\infty
e^{-z'^2} dz'\right),
\end{equation}
where $\Sigma_{CR} \simeq 100 \mbox{ g\ cm}^{-2}$ is the stopping
depth of cosmic rays and $\Sigma_0$ is the surface density of the
disk. We adopt a value for $\Sigma_{0}/\Sigma_{CR} = 30$, consistent
with FS03 and roughly twice the minimum mass solar nebula value at 1
AU. Note that FS03 incorrectly identify $\eta_0$ as the midplane
resistivity, rather than $\eta_{\rm mid} = \eta_0 \exp\left( \Sigma_0/4
\Sigma_{CR}\right)$, though this error does not carry through their
calculations of the magnetic Reynolds number (J. Stone 2006, private
communication).

\subsection{Initial and Boundary Conditions}
We initialize the disk in hydrostatic equilibrium in the vertical
direction, leading to a density profile $\rho(z) = \rho_0
\exp(-z^2/H^2)$, where $H = \sqrt{2c_s^2 / \Omega^2}$ is the scale
height of the disk. The initial magnetic field strength is
parameterized by the maximum plasma $\beta = 2\mu_0
P_g/B_0^2=400$. The field geometry is a zero-net flux vertical field
$B_z(x) = B_0 \sin(2\pi x/L_x)$. We measure length in units of the
disk scale height, $H$, time in units of $\Omega^{-1}$ (although we
plot quantities in terms of the orbital period $\torb = 2 \pi /
\Omega$), and set the vacuum permeability $\mu_0 = 1$. Our
computational domain has size $(x \times y \times z) = 1 \times 4
\times 4$, with resolutions for the models given in
Table~\ref{run_params}.

In these units, all simulations have a sound speed $c_s = 7.071 \times
10^{-4}$, rotation rate $\Omega = 10^{-3}$, midplane density $\rho_0 =
1$, and random perturbations in $\mathbf{u}$ with amplitude $1 \times
10^{-6}$ to seed the MRI. In order to vary the size of the dead zone,
we vary $\eta_0$ in equation~\ref{resistivity}, which in turn varies
the midplane resistivity. We parameterize the size of the dead zone
with the magnetic Reynolds number at the midplane $Re_{M} = c_s^2 /
(\eta_{\rm mid} \Omega)$.

Our boundary conditions are periodic in $y$, shear periodic in $x$
\citep{HGB95}, and periodic in $z$. While the last is not rigorously
accurate for a stratified disk such as ours, previous work
\citep{SHGB96} as well as our own direct comparison to vertical field
boundary conditions \citep{BNST95} suggest that it does not make a
significant difference for the density perturbations driven by the
MRI.

\subsection{Validity of Zero-mass test particles}
We monitor the gravitational force on a zero-mass test particle fixed
at the center of the grid. This can be directly translated into a
torque for a particle that does not move significantly in radius over
the course of the simulation.  While fixing the position of the planet
over the $\sim 100 t_{orb}$ duration of the simulation is an
approximation, Figure 6 of \citet{N05} shows that zero-mass objects
vary their semi-major axes by no more than about $10\%$ over a similar
time span.

If we make the assumption that Type I and turbulent migration are
separable, then our analysis will apply more generally to objects of
finite mass. However, even if these effects do couple, there exist a
class of objects for which our results are valid. Here, we derive a
range of masses for objects large enough to neglect gas drag and still
small enough to cause negligible density waves.

Stokes' drag law states that the stopping time of an object larger
than the mean free path of the gas it is moving through is
\begin{equation}
\label{stokes}
\tau_f = \frac{2 \rho_s a_s^2}{9 \mu},
\end{equation}
where $\rho_s$ and $a_s$ are the density and radius of the solid
object and $\mu$ is the dynamic viscosity of the gas
\citep[e.g.][]{W77}.

In order to determine the mass range in which an object can have both
drag force and gravitational back reaction on the disk neglected (ie,
a massless particle on a fixed Keplerian orbit), we consider the range
of sizes a body with solid density roughly that of Earth, $\rho_s
\simeq 3 \mbox{ g\ cm}^{-3}$, will have both drag times and Type I
migration times on the order of the lifetime of a protoplanetary disk,
taken here to be $10^7 \mbox{ yr}$. The dynamic viscosity of a pure
hydrogen gas is $\mu = 5.7 \times 10^{-5} T^{-1/2} \mathrm{ g\
cm^{-1}\ s^{-1}}$ \citep{Lang80}. Using an $\alpha$-disk with $\alpha
= 10^{-2}$ and $\dot{M} = 10^{-7} \mbox{ M}_\odot \mbox{ yr}^{-1}$,
\citet{IN06} give a midplane value of $T \simeq 100 \mbox{ K}$. The
lower limit, $\tau_f > 10^7 \mbox{ yr}$ gives a radius $a_s \sim 5.1
\times 10^4 \mbox{ cm}$ which for a spherical object of the above
density yields $m \gtrsim 4 \times 10^{14} \mbox{ g}$. Type I
migration times scale inversely with mass, $\tau_{\mbox{I}} \sim 8
\times 10^5 (m/\mbox{M}_\oplus)^{-1} \mbox{ yr}$ \citep{TTW02}, and so the
upper mass limit is $m_{max} \lesssim 10^{26} \mbox{ g}$. This is a
broad range of protoplanet masses, $10^{14} \mbox{ g} \lesssim m
\lesssim 10^{26} \mbox{ g}$, and suggests our results are relevant
even if Type I and turbulent migration are coupled for more massive
objects.

\section{Plasma Dynamics and Dead Zone Parameters} \label{sec:plasma}
\begin{figure}
\plotone{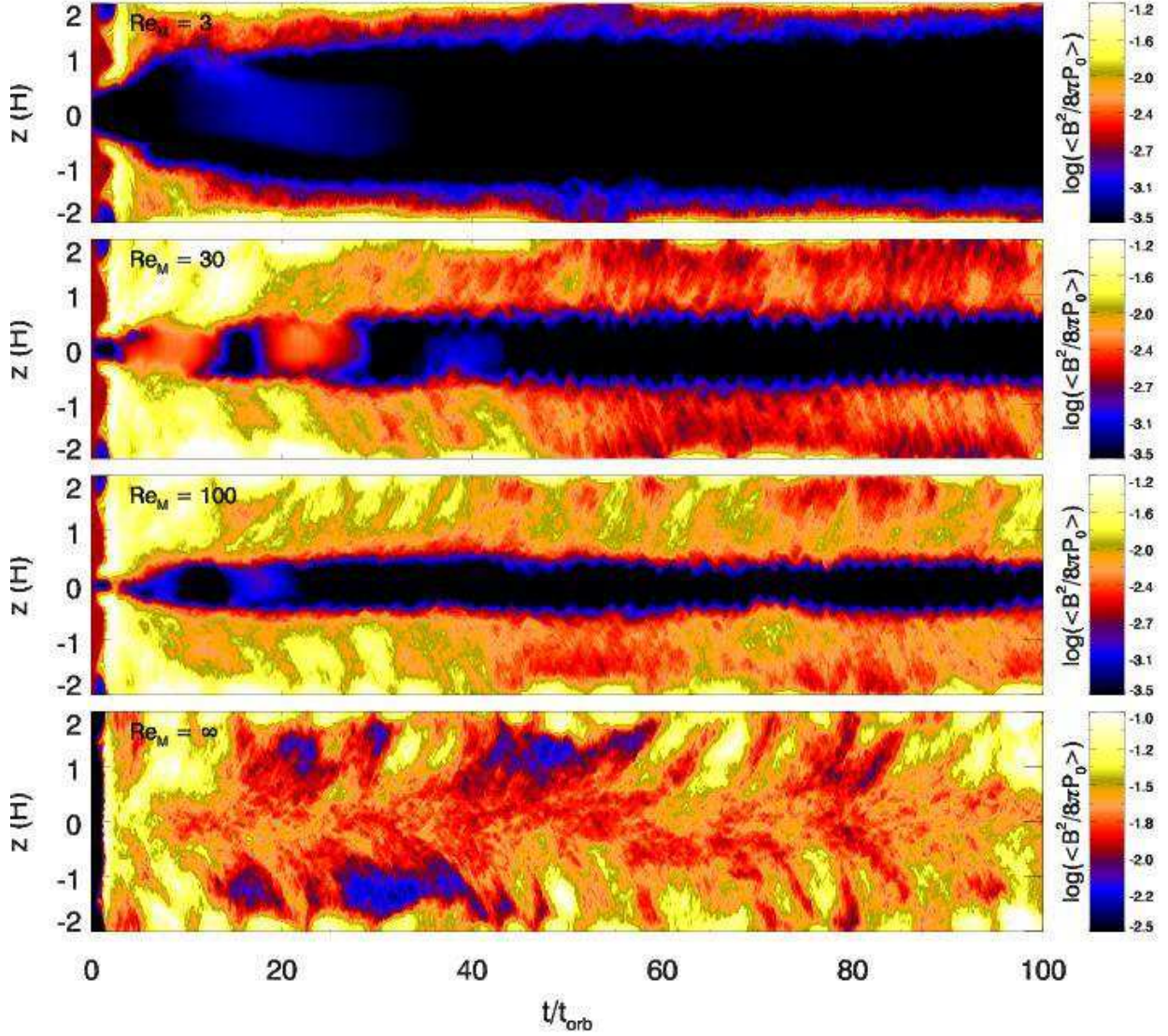}
\caption{
\label{xyaver_med_res}
Space-time diagrams of magnetic energy averaged over horizontal
($x-y$) planes for runs with (from bottom to top) $Re_M = \infty$,
100, 30, and 3. The dead zone is apparent in all models with finite
$Re_M$. Note that the color scale is shifted to higher values for the
bottom panel}
\end{figure}

In our models with finite $Re_M$, magnetic energy rapidly grows near
the midplane after a few orbits, due to the cooperation of magnetic
reconnection and shear, in agreement with the results of
\citet{FS03}. By 10 orbits, MRI turbulence sets in above and below the
midplane, forming active layers surrounding a quiescent (but not
motionless) dead zone that persists to the end of the simulation
(Figure~\ref{xyaver_med_res}). We define the dead zone to be the
region where the horizontally averaged Maxwell stress $-\langle B_x
B_y\rangle_{xy}$, has fallen by an order of magnitude from its average
value at $\pm 1.5 H$.  (We use unadorned angle brackets to denote
volume averages, while spatial averages over lower dimension and
temporal averages have identifying subscripts.)  Using $1.5 H$ as our
baseline, rather than the grid boundary at $2 H$, avoids contamination
by the slightly elevated value of the Maxwell stress that we see at
the boundary.  We have confirmed this effect is a result of the
periodic boundary conditions adopted with simulations using other
boundary conditions. We calculate the total surface density in
the active zones $\Sigma_A$ and compare that with the surface density
of the dead zone $\Sigma_D$ in Table~\ref{run_params}.

\begin{figure}
\plotone{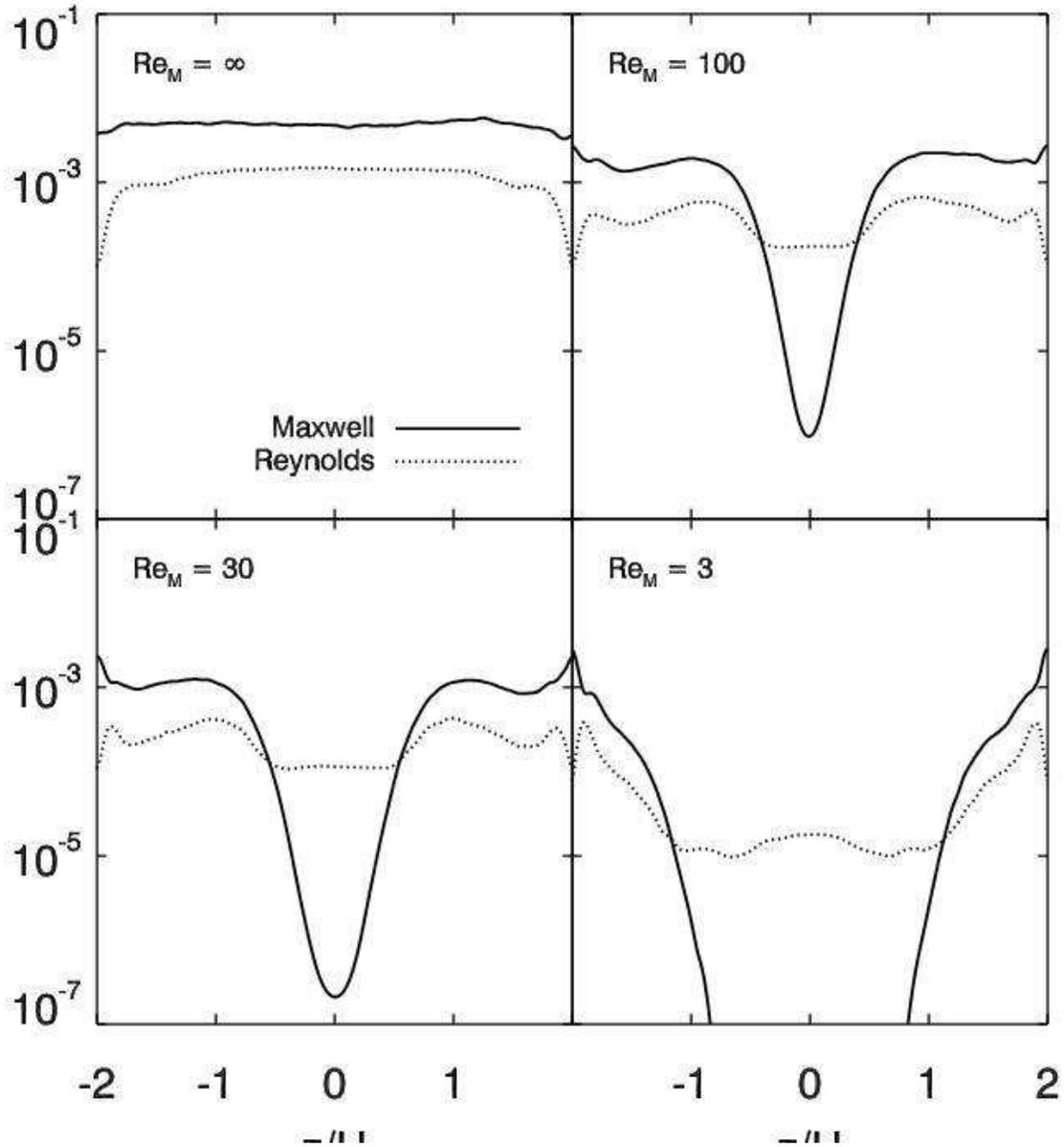}
\caption{
\label{alpha_vs_z}
Vertical profiles of Reynolds and Maxwell stresses for medium
  resolution runs with four different values of magnetic Reynolds
  numbers $Re_M$. Note that the Reynolds stress remains nearly
  constant across the dead zone, even as the dead zone depth
  increases.}
\end{figure}

In a shearing box, symmetry restricts the net transfer of angular
momentum, and thus the accretion rate, to zero. In order to measure
the ability of the turbulence to drive accretion, we measure the
Maxwell stress, as well as the Reynolds stress $\langle \rho u_x
u_y\rangle$.  These quantities measure the radial transport of angular
momentum by the turbulence \citep[e.g.][]{B98}.
Figure~\ref{alpha_vs_z} shows the turbulent stresses normalized by the
initial midplane pressure for each of the medium resolution runs. We
see Reynolds stresses significantly exceeding Maxwell stresses in the
midplane for all of our dead zone models. FS03 demonstrated that these
correlated motions are caused by momentum flux from turbulent
overshoot at the boundary layer between dead and active regions.

Our models confirm the finding of FS03 that a significant Reynolds
stress remains within the dead zone, even as the size and depth of the
dead zone, as defined by the dropping Maxwell stress,
increase. Because the criteria used by FS03 to define the dead zone
width are not entirely clear, it is difficult to make detailed
comparisons between our results and theirs. However, applying our
definition of a dead zone to Figure~3 in FS03, which shows Maxwell and
Reynolds stresses for a model with $Re_M = 100$, the dead zone extends
from $|z|<0.5$. In our comparable model, the dead zone is located
between $-0.35 < z < 0.38$. Given that FS03 report that their $Re_M =
10$ model was unable to sustain MRI turbulence while we were able to
run a self-sustaining $Re_M = 3$ model, it is possible that the
difference in dead zone widths is a result of the different
dissipative properties of our respective codes.

The $Re_M = 3$ case demands special attention. Its vertical
distribution of stresses is notably different than those of the other
non-ideal runs. First, the Reynolds stress at the midplane is
substantially less than in the $Re_M = 30$ and $100$ cases, which
differ by a factor of a few. Second, the Maxwell stress appears to
reach a minimum below $10^{-7}$ and remain constant over $|z|
\lesssim 0.5$, though it occasionally becomes slightly negative and is
excluded on the log plot. This behavior occurs in both 32R3 and 64R3
models, as well as for a run in double precision. We note that
Figure~5 of \citet{TSD06} shows similar behavior (though not the
negative $-B_x B_y$) in dead zone models computed with the Zeus code.

\subsection{Resolution Study}
\begin{figure}
\begin{center}
\includegraphics[height=3.5in]{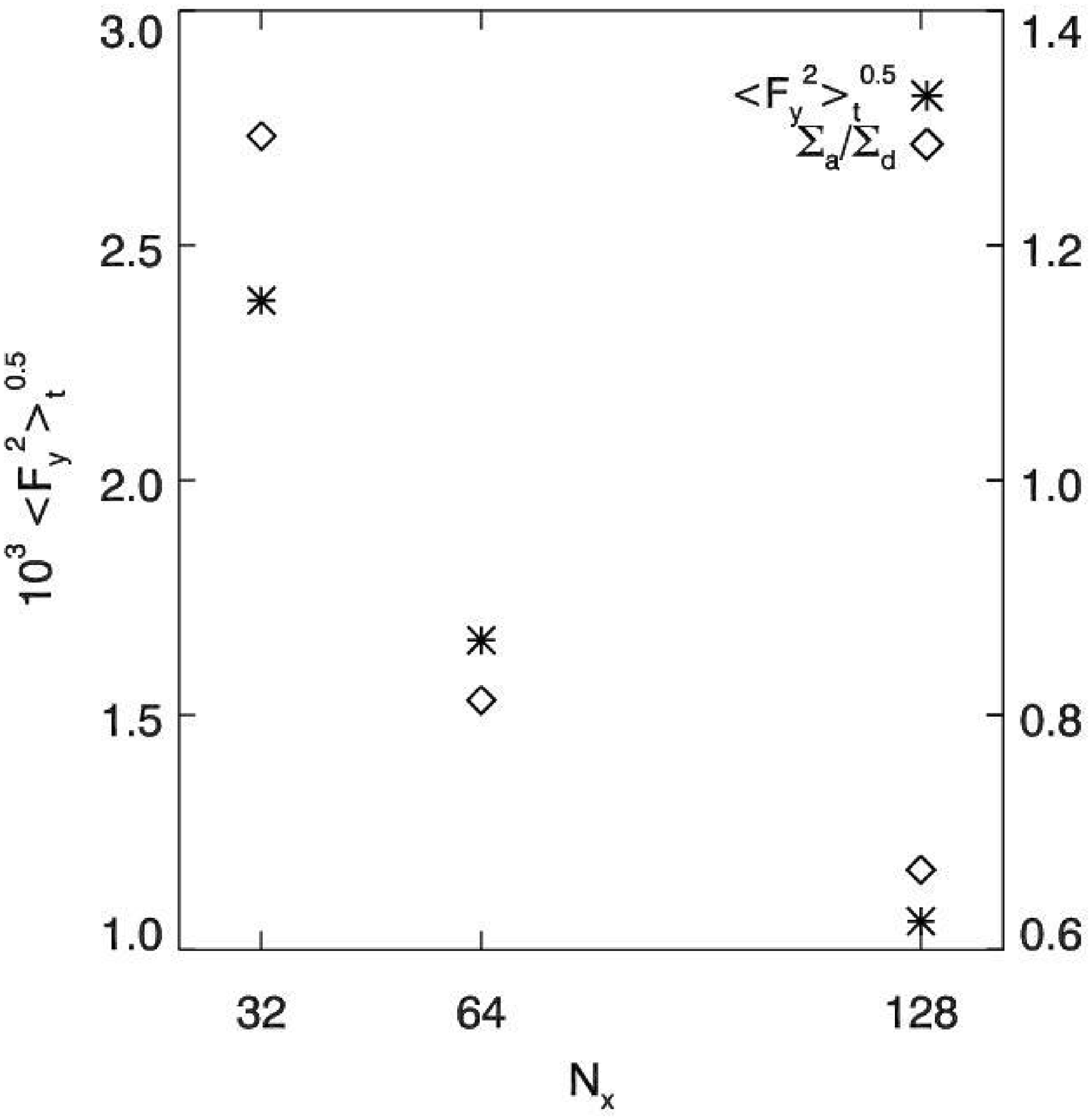}\\
\includegraphics[height=3.5in]{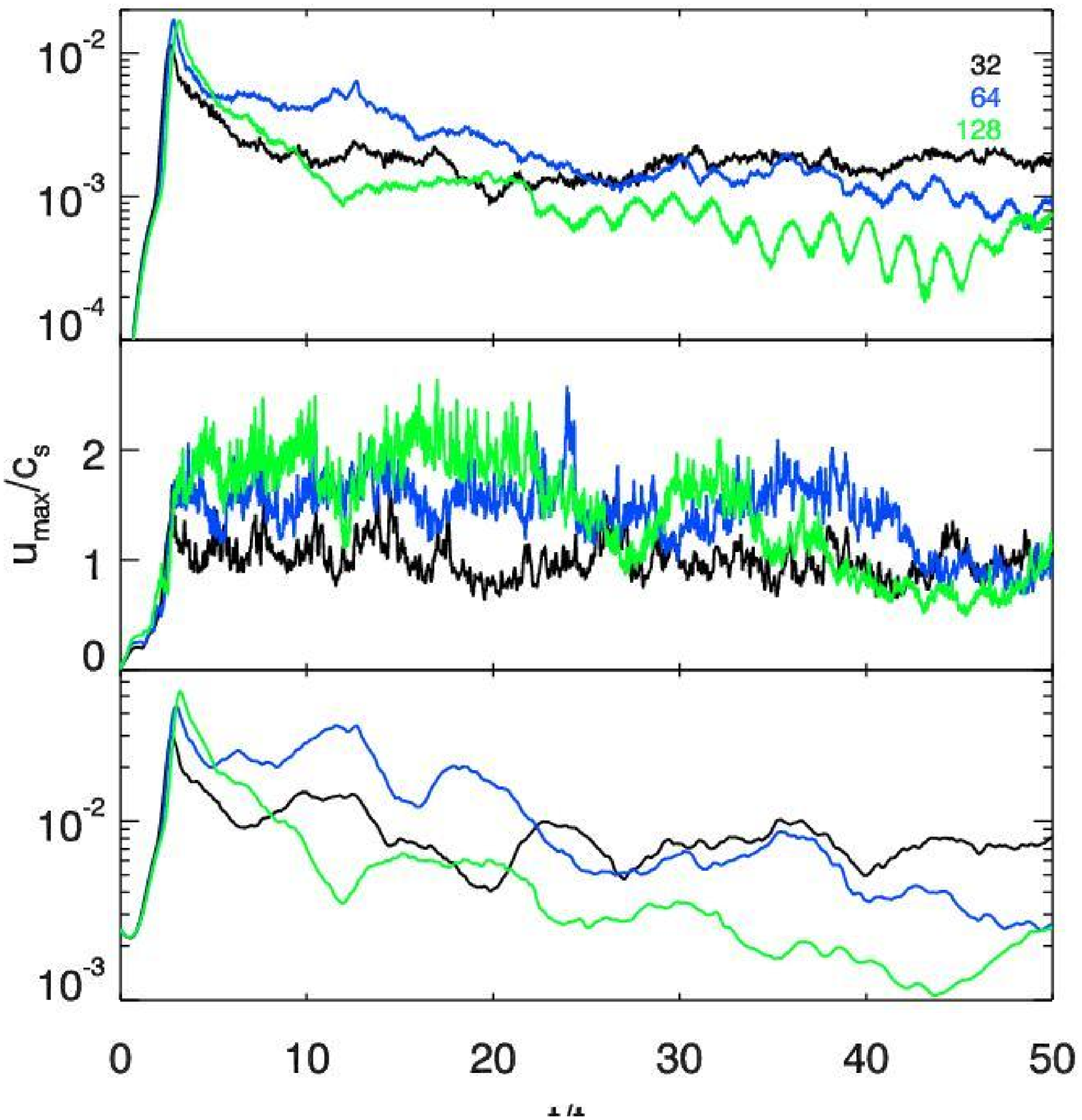}
\end{center}
\caption{
\label{res_study} A resolution study for a $Re_M = 30$ model. Resolutions corresponding to $dx =
3.125\times10^{-2}$, $1.563\times10^{-2}$, and $7.813\times10^{-3}$ are
labeled by $N_x = 32$, 64, and 128. Variables shown include \emph{(a)} the
temporal RMS azimuthal gravitational force $\langle F_y^2
\rangle_t^{0.5}$ ({\em stars}) and the active-to-dead zone column
density ratio $\Sigma_a/\Sigma_d$ ({\em diamonds}) versus resolution;
and \emph{(b)} volume averages of the effective dimensionless
viscosity $\alpha = (\langle \rho u_x u_y\rangle - \langle B_x B_y
\rangle)/P_0$ \emph{(top)}, RMS Mach number \emph{(middle)}, and
magnetic energy normalized to initial midplane pressure $P_0$
\emph{(bottom)}, as a function of time in orbital units.}
\end{figure}

In order to study the effects of numerical resolution on our results,
we ran our $Re_M = 30$ model at three resolutions, and all others at
two. An overall picture of the MRI is given by its saturation energy
or average viscous $\alpha$ value, both of which are shown in
Figure~\ref{res_study}. A more relevant quantity for our purposes is
the temporal root-mean-square (RMS) azimuthal gravitational force
$(\langle F_y^2 \rangle_{t})^{0.5}$ (i.e., torque; see
\S~\ref{sec:torques}). This quantity is directly responsible for
orbital migration and is given in
Table~\ref{run_params}. Figure~\ref{res_study}a shows the RMS torque
and dead zone column density ratio $\Sigma_a / \Sigma_d$ as a function
of resolution. While both are resolution dependent, it appears that
their values are converging between the highest two resolution
runs. The dimensionless effective viscosity $\alpha = (\langle \rho
u_x u_y\rangle - \langle B_x B_y\rangle)/P_0$ and maximum Mach number
are given in the upper two panels of Figure~\ref{res_study}b. Both
quantities reflect the well-known result that higher resolution MRI
runs lead to lower turbulent stresses \citep{FP07}.

The magnetic energy for the high resolution run drops by nearly a
factor of $7$ between orbits $30$ and $50$, similar to the simulations
of \citet{SHGB96}. Decreases in magnetic energy at similar times and
on similar timescales occur at all resolutions, although not with such
large magnitudes. In the lower resolution models run to longer times,
these drops eventually stop occurring, typically by orbit $100$ and
the model maintains a relatively constant magnetic energy level. The
timescale for such an adjustment is curiously long; we hope to
examine it more closely in future work. For our purposes here, it is
sufficient to note that while resolution effects are present, the
\emph{duration} of the simulation may be a greater factor in obtaining
robust statistical results \citep*[see][for a discussion]{WBH03}.

\section{Migration Torques} \label{sec:torques}
Turbulent overdensities in a protoplanetary disk exert torques on a
protoplanet at radius $R$ with strength 
\begin{equation}
\mathbf{\Gamma} = d\mathbf{J}/dt = \mathbf{R \times F}.
\end{equation}
The sum of gravitational forces from all turbulent density
perturbations is
\begin{equation}
\label{f_y_eqn}
\mathbf{F} = \sum_i \frac{G \rho_i \Delta V}{r_i^2}\mathbf{\hat{r}_i},
\end{equation}
where $\mathbf{r_i}$ is the distance to each gas zone with density
$\rho_i$, $\Delta V = \Delta x \Delta y \Delta z$ is the zone volume
and the sum is taken over all zones in the computational domain. In
particular, orbital migration is caused by a change in angular
momentum $\mathbf{J}$, which for a circular orbit in a spherically
symmetric potential is simply $J_z$. Following \citet{NP04}, we
consider only a protoplanet at constant $R$---the center of our
shearing box, and thus $dJ_z/dt = \Gamma_z \propto F_{y}$. In the
following analysis, we consider only the properties of the scalar
$F_{y}$ and thus only track migration on circular orbits. We scale
$F_y$ in units of $2 \pi G \Sigma$, which is the force per unit mass
felt by a particle suspended a small distance above the center of a
disk with constant $\Sigma$. We sample this quantity every $100$
timesteps throughout the calculation. While the timestep is
dynamically determined and thus fluctuates over the simulation, it
maintains an average value of $\sim 10^{-4} \mathrm{\torb}$ with an
RMS at least an order of magnitude smaller, giving a sampling rate of
roughly $\Delta t \sim 10^{-2} \mathrm{\torb}$.

\begin{figure}
\plotone{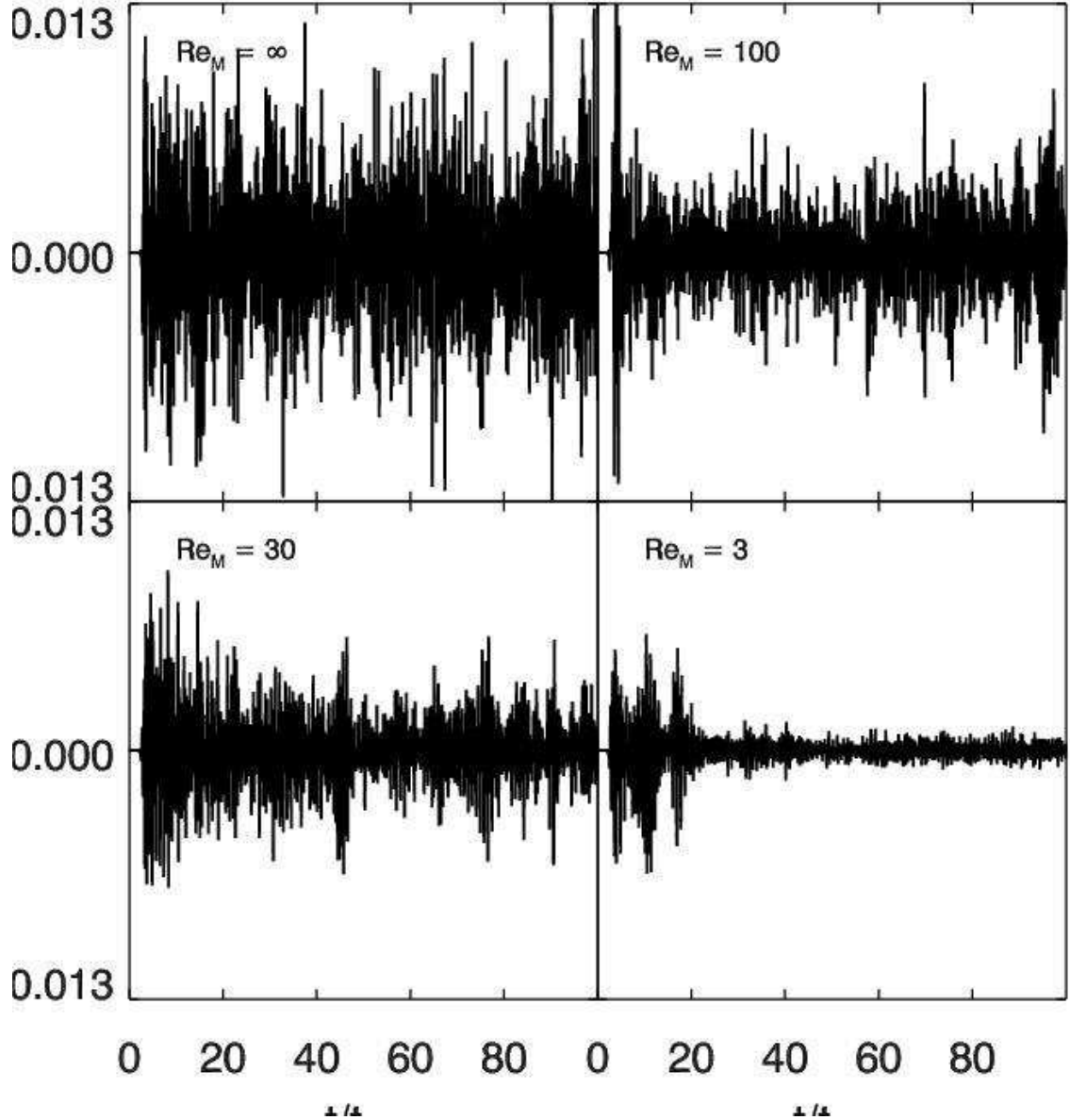}
\caption{
\label{fy_vs_t}
Time series of azimuthal gravitational force $F_y$ in $64 \times
256^2$ resolution runs. The force is scaled in units of $2 \pi G
\Sigma$, the vertical restoring force near the midplane.  }
\end{figure}

Figure~\ref{fy_vs_t} shows $F_y$ as a function of time for
simulations with varying values of $Re_M$. The RMS value of the force
declines with increasing dead zone size. This decrease occurs because
it is the MRI in the active layers that drives the density
fluctuations, which are in turn responsible for the azimuthal
force. For the remainder of the analysis, we consider only torques for
times $t > 25 \torb$ in order to avoid any of the transient features from the
onset of the MRI evident in Figure~\ref{fy_vs_t}. We will refer to the
value of $F_y$ as ``torque,'' although it is formally a force per unit
mass.

Two critical assumptions of the Fokker-Planck model for planetary
migration of JGM06 are that the turbulent torques have temporal
stationarity and finite correlation time. We test each of these
assumptions for the torque distribution in our simulations.

\subsection{Torque Distribution}
\begin{figure}
\plotone{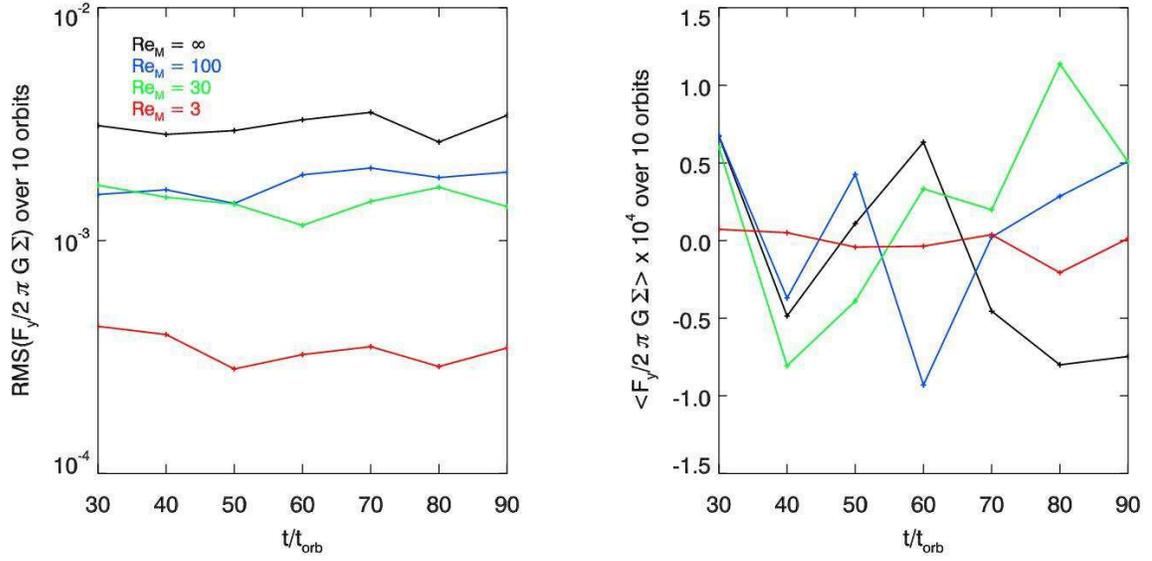}
\caption{
\label{rms_10_orbits}
RMS (left) and mean values (right) of the torque $F_y$ for seven
periods each of length $10 \torb$, labeled by the interval
midpoint. The $10 \torb$ means show weak fluctuations around the
global time standard deviation, as much as a factor of $\sim 0.2$ in
the $Re_M = 3$ case, and a factor of a few in all cases.  }
\end{figure}

In order to determine time stationarity, we need the distribution of
torques over various fixed time intervals. We can then ask if the
samples in each interval are consistent with being drawn from the same
distribution as the other intervals. We separate the torque time
series into seven $10 \torb$ blocks from $25-100
\torb$. Figure~\ref{rms_10_orbits} shows that the mean of each
interval remains near zero over time, and the RMS remains nearly
constant. JGM06 require that the torque distribution be such that
$\overline{\delta \Gamma(t,J)^2}$ has no time dependence, and this
plot demonstrates this for our study. However, the means over each $10
\torb$ bin vary about zero with amplitudes a factor of $\lesssim 0.2$
times the standard deviation over the entire $75 \torb$ sample,
suggesting that the underlying distribution of torques does vary
slightly over this short an interval. This does not affect the
stochastic migration presented here, as such an interval is much
shorter than a typical diffusion time in the JGM model.

\begin{figure}
\plotone{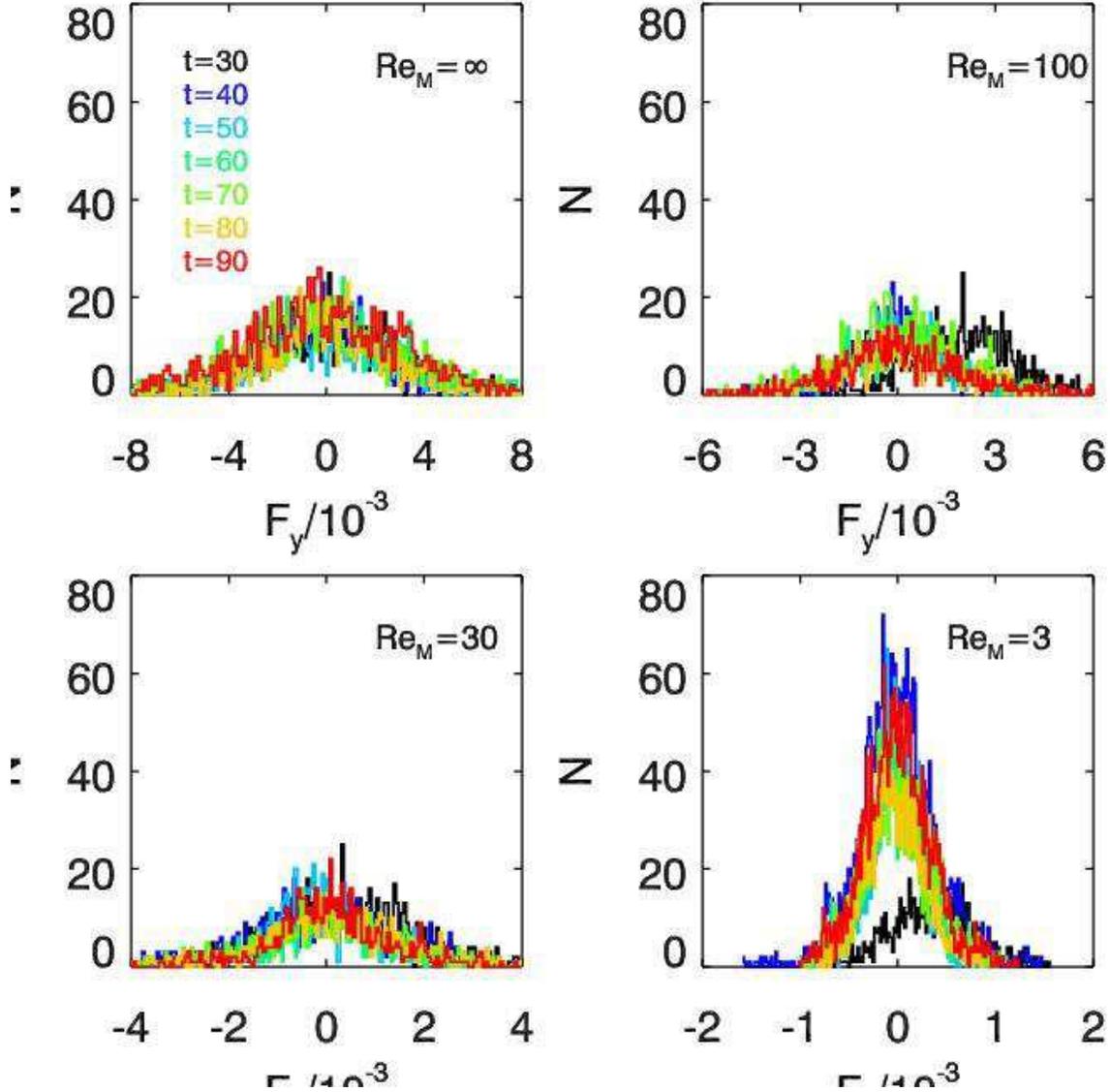}
\caption{
\label{torque_histogram}
Distributions of torque exerted on protoplanets in models with various
$Re_M$. Each panel shows the distributions for seven periods, each of
length $10 \torb$ (from $25 \torb$ to $95 \torb$), labeled by the
interval midpoint. Note the decreasing range along the $x$-axis from
model to model, indicating a narrower range of torque amplitudes for
larger dead zone models.  The distributions appear time stationary
after the initial 10 orbits.  }
\end{figure}

Additionally, we are interested in the particular distribution of
these torques: the central limit theorem assures us that the
cumulative effect of any random process limits to a Gaussian
distribution, but can we expect to have a coherent Gaussian over a
given (short) period, given the unknown distribution of turbulent
forces?  Figure~\ref{torque_histogram} shows the distributions of
torque over $10$ orbit periods for each of the dead zone widths. Each
period represents $\sim 10^4$ samples. There is a significant change
in the distribution for the $Re_M = 3$ case, although it is consistent
with the trend of decreasing standard deviation with decreasing
$Re_M$. Qualitatively, even over such short periods, the torques
appear Gaussian distributed, in the sense that they are centrally
concentrated on zero and roughly symmetric
(figure~\ref{torque_histogram}). We quantify this by computing the skew
\begin{equation}
S = \frac{1}{N} \sum_{j=1}^N \left[ \frac{x_j - \overline{x}}{\sigma} \right]^3,
\end{equation}
and kurtosis
\begin{equation}
K = \frac{1}{N} \sum_{j=1}^N \left[ \frac{x_j - \overline{x}}{\sigma}
  \right]^4 - 3,
\end{equation}
where $\sigma$ is the standard deviation of the $N$ samples, given in
tables~\ref{skew} and~\ref{kurtosis} respectively. A normal
distribution has skew and kurtosis both equal to zero, while an
exponential distribution has kurtosis of three. These tables show that
over 10 orbits, the distribution varies from normal
significantly---the variance from normal is given using the standard
$\sigma_{var} = \sqrt{24/N}$ for kurtosis \citep{numrec}. This is
directly related to the chaotic nature of the MRI: small deviations in
initial conditions present at the start of each $10 \torb$ block will
cause significant deviations \citep{WBH03}, which we observe as
differences in distribution. We note here that this analysis does not
affect the JGM assumptions--their model does not require a Gaussian
distribution (shown above)--it merely underscores the short-term
properties of the MRI.

\subsection{Fourier Analysis}
\label{fourier}
\begin{figure}
\plotone{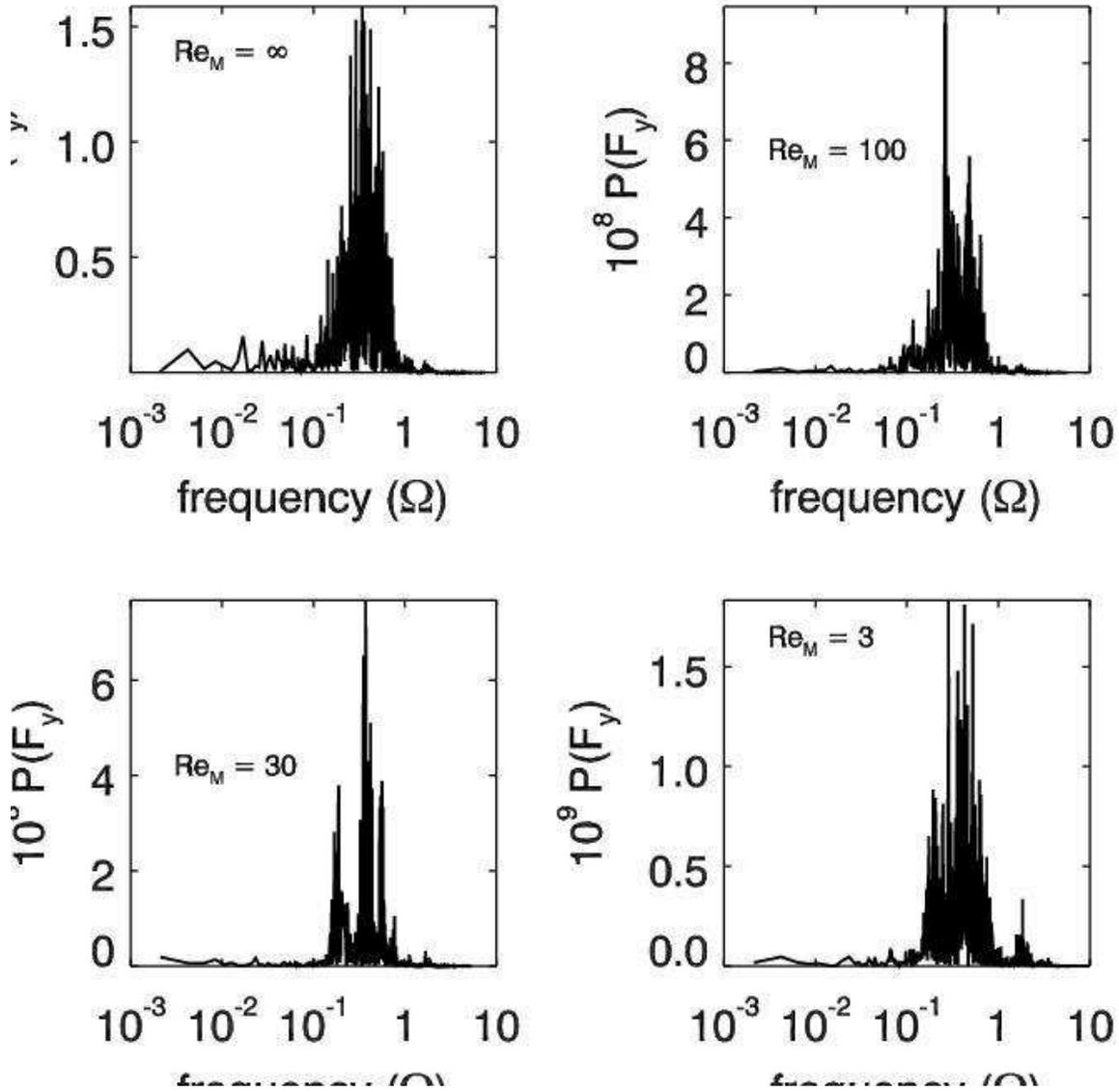}
\caption{
\label{mig_powerspectra}
Power spectra of torque fluctuations $F_y$ in four medium resolution
models with increasing dead zone width. Note the decreasing value of
the vertical scale from model to model. Very little power is seen at
low frequencies.}
\end{figure}

Figure~\ref{mig_powerspectra} shows the temporal power spectra of the
torque for runs with varying dead zone widths. Increasing dead zone
width has little effect on the shape of the spectrum, changing only
the total power. This suggests that the turbulence produces a similar
spectrum of density fluctuations, regardless of the width of the dead
zone, as we would expect, since the underlying instability is not
significantly altered by Ohmic resistivity \citep{J96,FSH00}.

\begin{figure}
\plotone{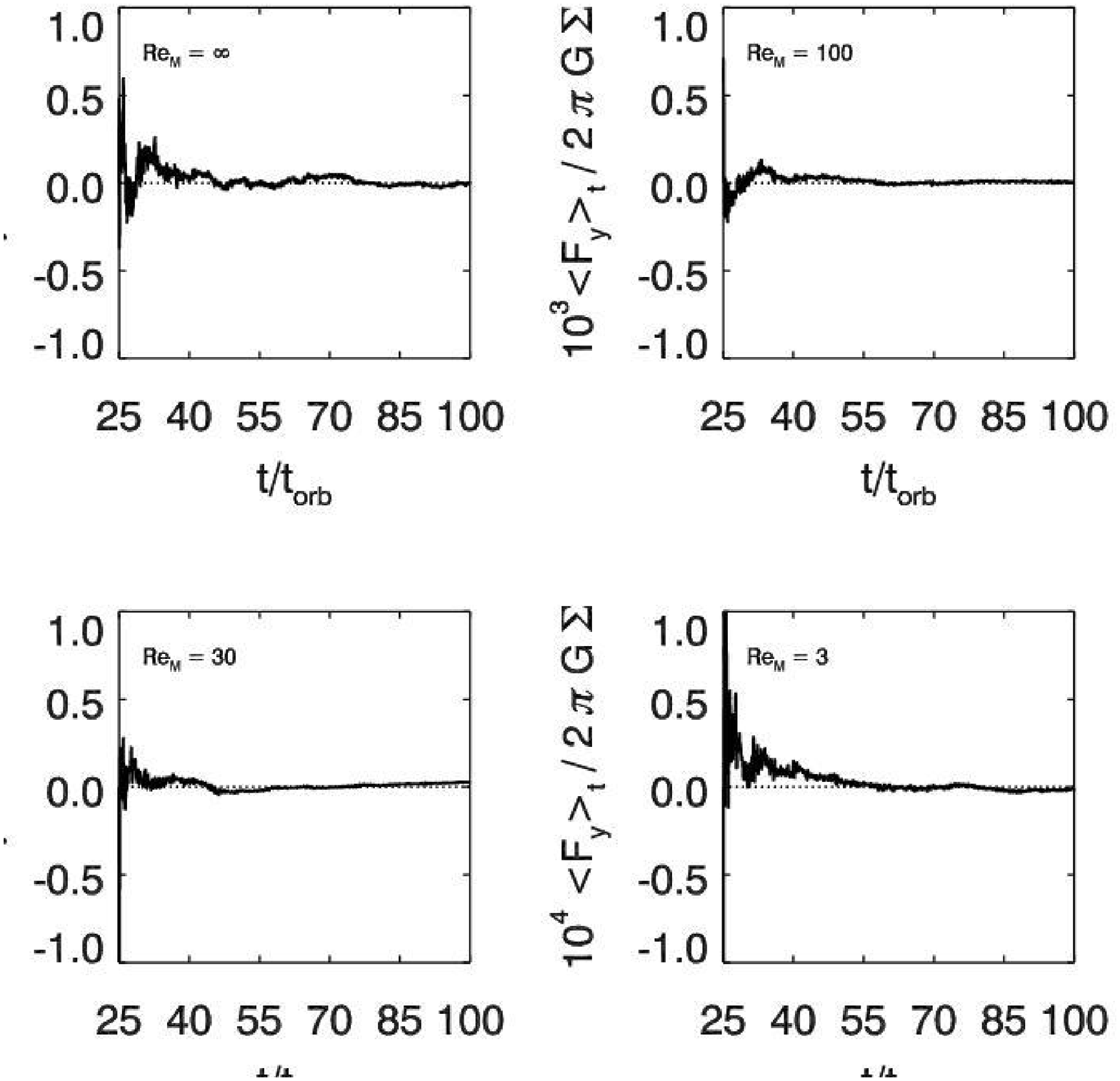}
\caption{
\label{mig_ravg}
Running time average of the torque $F_y$ in the four medium resolution
models with increasing dead zone size, starting at $t = 25 t_{\rm
orb}$ to eliminate effects of the initial transient. The dotted line
marks the zero point. The ordinate is a factor of ten smaller for the
$Re_M = 3$ case than for the others. In all cases, the running average
approaches zero at late times. }
\end{figure}

The low frequency power reported by \citet{N05} is absent in our
models, done in the local frame.  This suggests that whatever imparts
a long-term memory to the turbulence in his simulations is related to
the global structure. This conclusion was also noted in \citet{NP04}
on the basis of running time averages for the migration torque in
their local simulations. They find that running averages of the
torques tend to zero, while for low mass planets in global
simulations, they do not. Our stratified but non-resistive models show
a similar convergence, which tends to smaller values at late times for
increasing dead zone width (Fig.~\ref{mig_ravg}).

As noted in \S~\ref{sec:torques}, our time series data is not evenly
spaced in time. To correct for this, we interpolate to an even time
grid for Fourier analysis. While this can be dangerous for
astronomical time series with large gaps \citep[eg.,][]{S82}, we have
confirmed that in our case, interpolation is essentially identical to
a Lomb Normalized Periodogram \citep{numrec}, which produces power
spectra (but not correlations) for unevenly sampled time series data.

\begin{figure}
\plotone{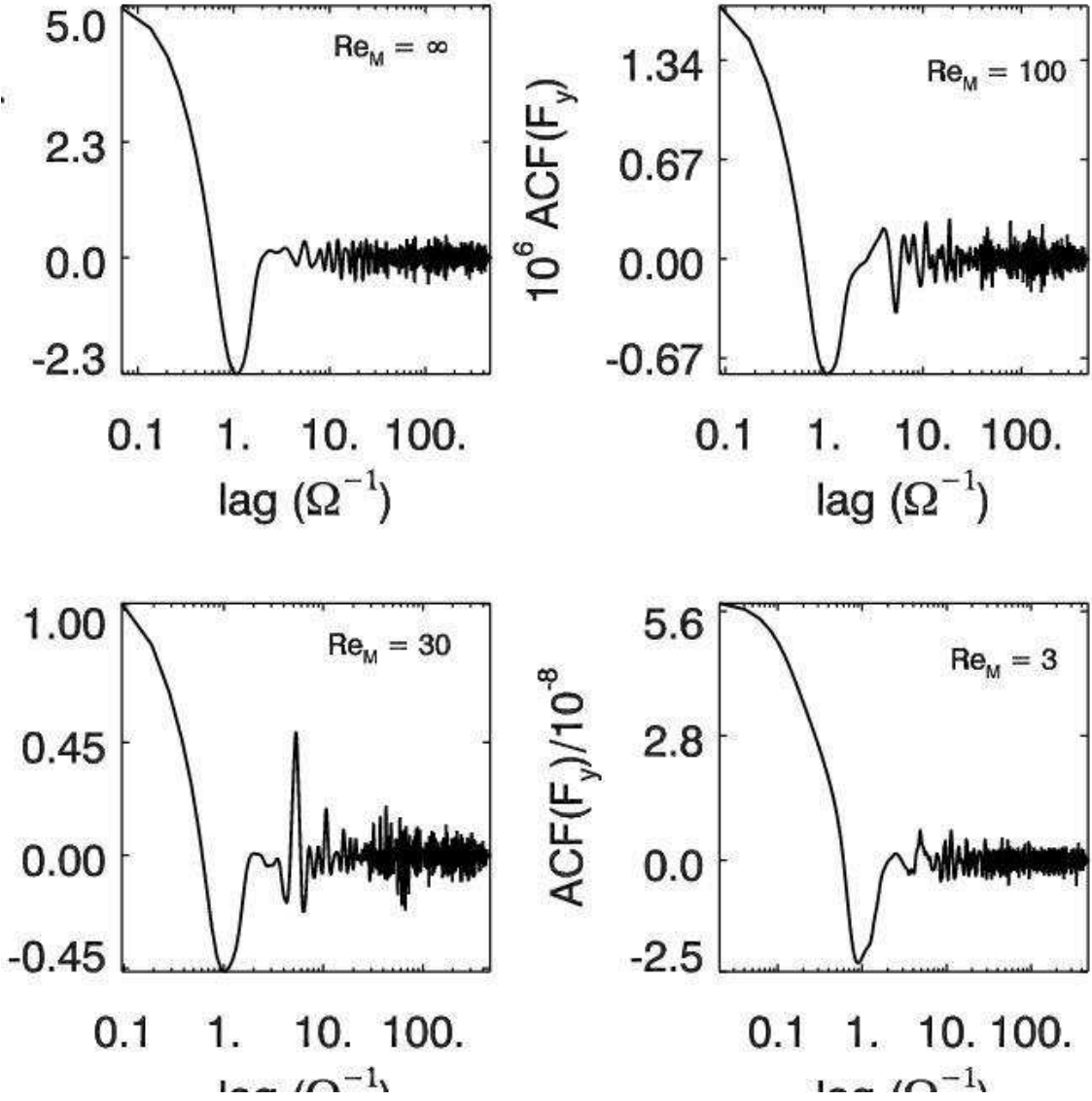}
\caption{
\label{mig_acf}
Autocorrelation functions of the torque ($F_y$) fluctuations in the
four medium resolution models with increasing dead zone width. All
models have similar correlation times, roughly defined as the point
where the curve first crosses zero with positive first derivative. }
\end{figure}

The existence of discrete peaks in the power spectrum indicates the
presence of characteristic timescales for the torquing of the
protoplanet by MRI turbulence. A more direct measure of this timescale
can be found with the autocorrelation function, which we calculate by
taking the inverse Fourier transform of the power
spectrum. Figure~\ref{mig_acf} shows autocorrelation functions for
each of the medium resolution runs. There is almost no correlation for
lags longer than about a dynamical time. Thus, in a local frame at
least, turbulent torques are correlated only for a short time,
typically $\sim 0.3 \torb$ (see Table~\ref{run_params}).

\subsection{Locality}
\begin{figure}
\plotone{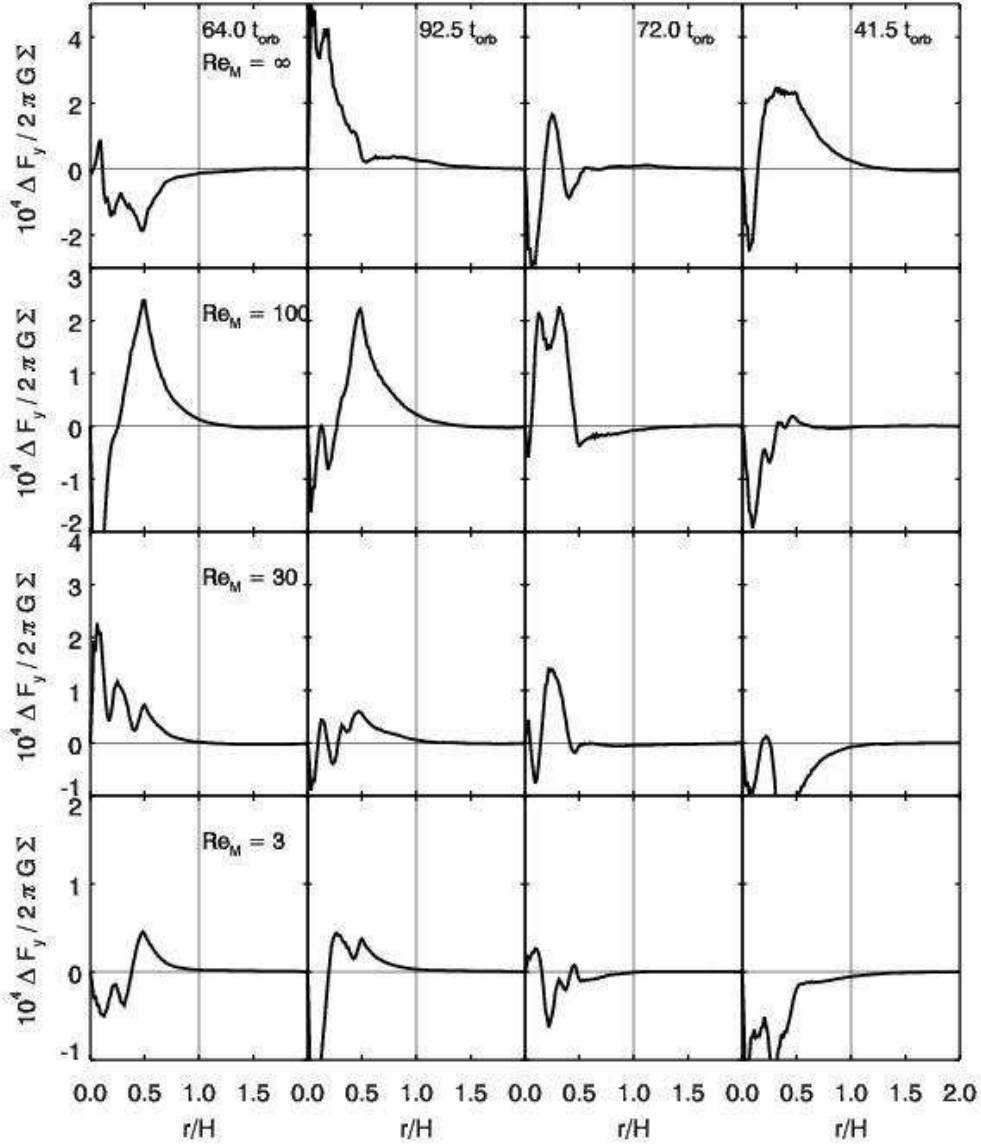}
\caption{
\label{mig_location}
Cumulative torque contributions $\delta F_y$ from spherical shells as
a function of radial distance $r$ from the box center shown at four
random times in medium resolution models with increasing dead zone
size (from top to bottom, $Re_M = \infty,100,30,3$). In all models,
the torque vanishes at or near $r \sim H$, the disk scale height. The
thin lines mark the zero point on the ordinate and $r = H$ on the
abscissa.}
\end{figure}

It is interesting to understand where the forces relevant to turbulent
migration come from. Are they dominated by local forces, or do
increasingly large perturbations from large scale motions dominate the
total force? Figure~\ref{mig_location} shows the total instantaneous
force contributed by concentric spherical shells centered on the
protoplanet at the box center, calculated as $\Delta F_y(r) = \sum_i
F_y(x_i,y_i,z_i)$, for $r^2 < x_i^2+y_i^2+z_i^2 < (r+dr)^2 $, at four
random times for each model. All panels show a sharp drop-off outside
of $r \simeq \mathrm{H}$, validating the local approach and
demonstrating that nearby perturbations tend to dominate over larger
turbulent structures.

However, the shearing box imposes a cutoff on the outer scale of MRI
turbulence that precludes very large eddies from forming.  In
particular, at scales large enough for the disk to act as a
two-dimensional flow, eddy formation may occur from the inverse
cascade of energy expected in two-dimensional flows. We cannot rule
out the possibility that such structures may affect protoplanetary
migration. JGM06 argue that large eddies will dominate the diffusion
coefficient in a turbulent flow because they have longer correlation
times. Our simulations do not address this, as we do not have
information about the correlation time as a function of
scale. However, we have run a model with twice the radial ($x$) extent
that shows a similar drop-off at a scale of $\mathrm{H}$ ensuring that
our result is not simply a numerical effect of the small radial size
of our box. In addition, we have some insight from the observation
that the correlation times differ very little through the sequence of
increasing dead zone sizes (see \S~\ref{fourier}). This suggests that
the total effect of the various size density structures does not
change, even as the various eddy sizes change as a result of the dead
zone. Future work will address this point.

\begin{figure}
\plotone{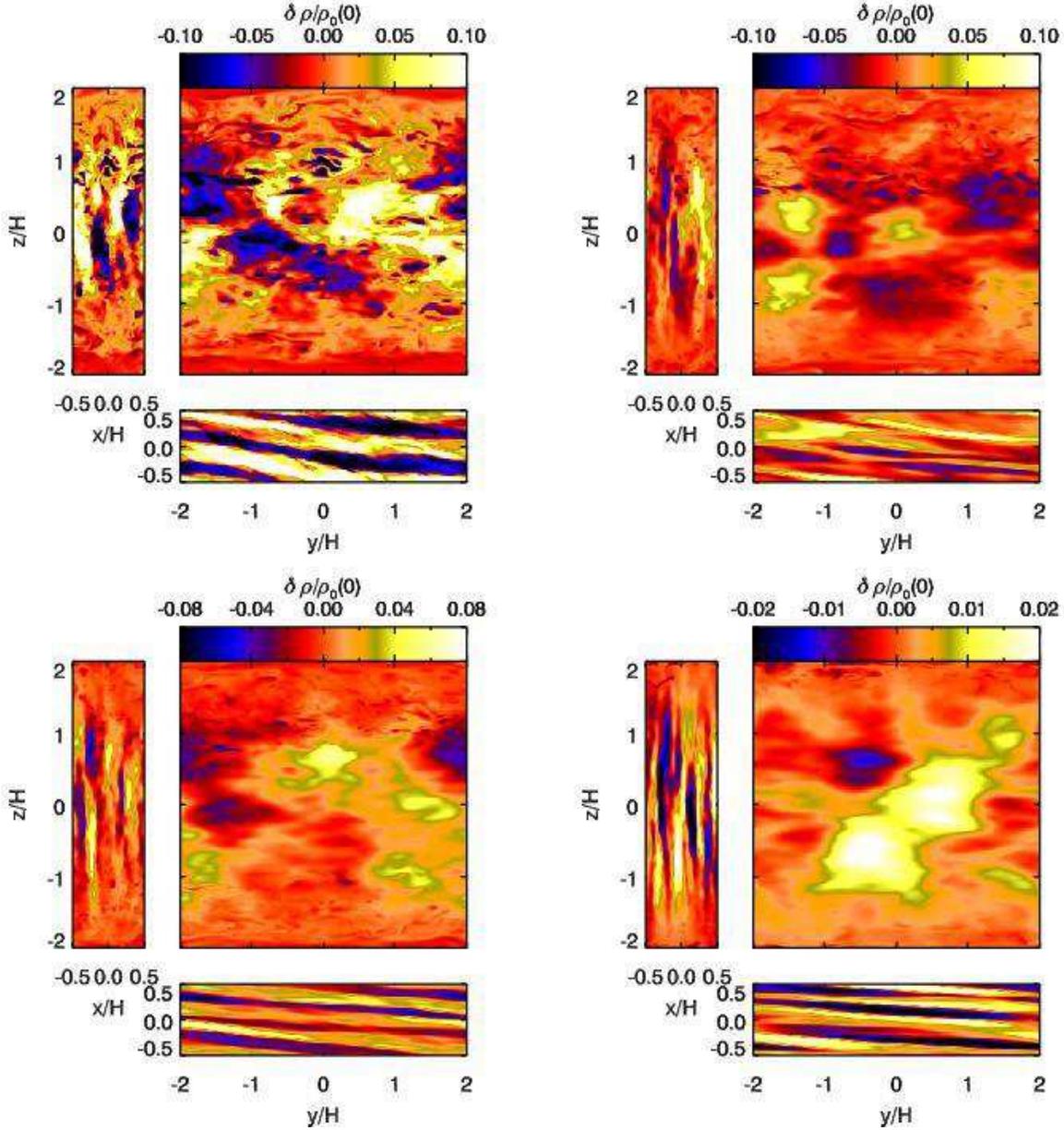}
\caption{
\label{img_drho_med_res}
Turbulent overdensities $\delta \rho = \rho - \rho_0$, where $\rho_0$
is the initial hydrostatic distribution, in three orthogonal planes
through the origin in each model with increasing dead zone width and
normalized by the initial midplane density, $\rho_0(0)$. $xz$ and $yx$
cuts surround the $yz$ cut. All sub-panels share a common color scale,
given by the colorbar over the $yz$ images. Clockwise from top left,
$Re = \infty, 100, 30,3$. As the dead zone width increases, a clear
transition from turbulent to wave-like behavior is visible at the
midplane.}
\end{figure}

Overdensities at the midplane dominate the torque not only in the
fully ionized model, but even in models with large dead zones, for two
reasons: obviously, they are closer to the protoplanet, and secondly,
less obviously, the higher total density at the midplane means that
even small perturbations there produce absolute overdensities higher
than in the turbulent active layers. Figure~\ref{img_drho_med_res}
shows images of the turbulent overdensities, $\rho - \rho_0(z)$ at $t
= 100 \torb$ for all values of $Re_M$. Subtracting off the initial
exponential profile $\rho_0(z)$ simply removes the large symmetric
component due to the unperturbed density distribution, allowing the
density fluctuations that contribute to the torque to be more easily
visualized. The scale bars show the general decline in density
perturbation amplitude. Two features are worth noting: the increasing
dominance of wave-like structures in these torque images, and the fact
that the torque drop-off does \emph{not} come from the MRI moving
further away from the planet. While that does happen, the strongest
density perturbations remain near the midplane, regardless of the
presence of the MRI at that location. The decrease in torque strength
for lower magnetic Reynolds numbers is due to the overall lessening of
turbulent energy because of both thinner active layers, and reduced
energy at saturation due to the higher Ohmic resistivity present even
in the active layers. For our ideal MHD runs, $\delta \rho$ is
consistent with the relation 
\begin{equation}
\frac{\left< \delta \rho^2 \right>^{0.5}}{\left< \rho
\right>} \simeq \frac{\left< \delta P_{mag}^2 \right>^{0.5}}{\left< P \right>}
\end{equation}
\citep{SITS04}, suggesting that their result also applies to
stratified disks. For our isothermal runs,$\left< P \right> = \left<
c_s^2 \rho \right> = c_s^2 \left< \rho \right>$ and the relation
reduces to $\left< \delta \rho^2 \right>^{0.5} \simeq \left< \delta
P_{mag}^2 \right>^{0.5}/c_s^2$. Table~\ref{sano_density_relation}
shows the relationship between volume averaged density and magnetic
pressure perturbations for our medium resolution models. As the dead
zone size increases so too do the density perturbations relative to
the variance of magnetic pressure. This is consistent with the
observation of density waves in the dead zone: the active zones drive
density perturbations in regions with negligible magnetic pressure,
and so volume averages should show just such a trend to higher ratios
of the two quantities.

The wave-like structures that we see in the simulations may be the
low-$m$ spiral density waves reported in earlier shearing box studies
\citep*{HGB96}. The wave pattern appears similar in all cases because
it is driven by the MRI. We switched off the Lorentz force in one of
our $Re_M = 30$ simulations, and allowed the evolution to continue for
about 6 orbits. With no MRI driving in the active layers, the
turbulence quickly dissipates. However, during the decay, the wave
pattern remains, although its amplitude drops by roughly a factor of 8
in those 6 orbits. This suggests that the wave pattern may be a
natural resonant state of either real disks or merely the shearing
box. Although the images reveal wave-like structure, the temporal
power spectra of torque do not differ significantly between fully
turbulent and dead zone models. Simulations of stochastic excitation
of solar $p$-modes \citep{NS01,SN01} tend to show strong temporal
signatures of the excited waves, while our power spectra do not show
any obvious wave signatures. We will investigate these structures
further in future work.

\subsection{Parameterization for Stochastic Models}
One of our objectives in this work is to provide parameters necessary
for refining the statistical treatment of migration by JGM06 and to
understand the predictions it makes when we include dead zones. The
extreme computational cost of three-dimensional MHD simulations
ensures that the fully global simulation including several scale
heights and a few tens of AU in radius evolved to the $\sim 10^5
\mathrm{\torb}$ required to follow planetary migration directly is
unlikely in the foreseeable future. Thus, a promising tool in the
study of planetary migration is a stochastic method in which one
considers the fates of a population of planets subject to random
torques from the disk turbulence. JGM06 have constructed an
advection-diffusion equation from the Fokker-Planck equation in which
standard Type I migration is modeled as advection and turbulent
torques are modeled as diffusion (note the important correction in
the erratum).  Understanding the long term evolution of a
population of planets in this model requires a measurement of the
diffusion coefficient due to turbulent torques in regions of disks
with varying resistivity profiles.

JGM06 derive this diffusion coefficient in terms of the
independent variable $J$, the orbital angular momentum of a
protoplanet at a given radius $r_p$, rather than the radius
itself. From a standard Fokker-Planck derivation, JGM06 define
the diffusion coefficient as
\begin{equation}
  \label{diffusion}
D(J) \equiv \int_{-\infty}^{\infty} \overline{\delta
  \Gamma(t-\tau/2,J) \delta \Gamma(t+\tau/2,J)} d\tau
  \simeq \overline{\delta \Gamma(t,J)^2} \tau_c,
\end{equation}
where $\tau_c$ is the correlation time, estimated here from the
autocorrelation function (\S~\ref{fourier}), and the overbars
represent ensemble averages. Direct calculation of the integral from
our time series for $F_{y} \propto \Gamma(t,J)$ does not yield
satisfactory results, probably because each simulation is only one
realization of the torque fluctuations, and $D(J)$ is well-defined
only for an ensemble of such realizations. 
In lieu of direct calculation, we make use of the rough equivalence
$D(J) \simeq \langle \delta \Gamma(t,J)^2 \rangle_{t} \tau_c$, where
we have substituted a time average for the ensemble average in the
original. 

\begin{figure}
\plotone{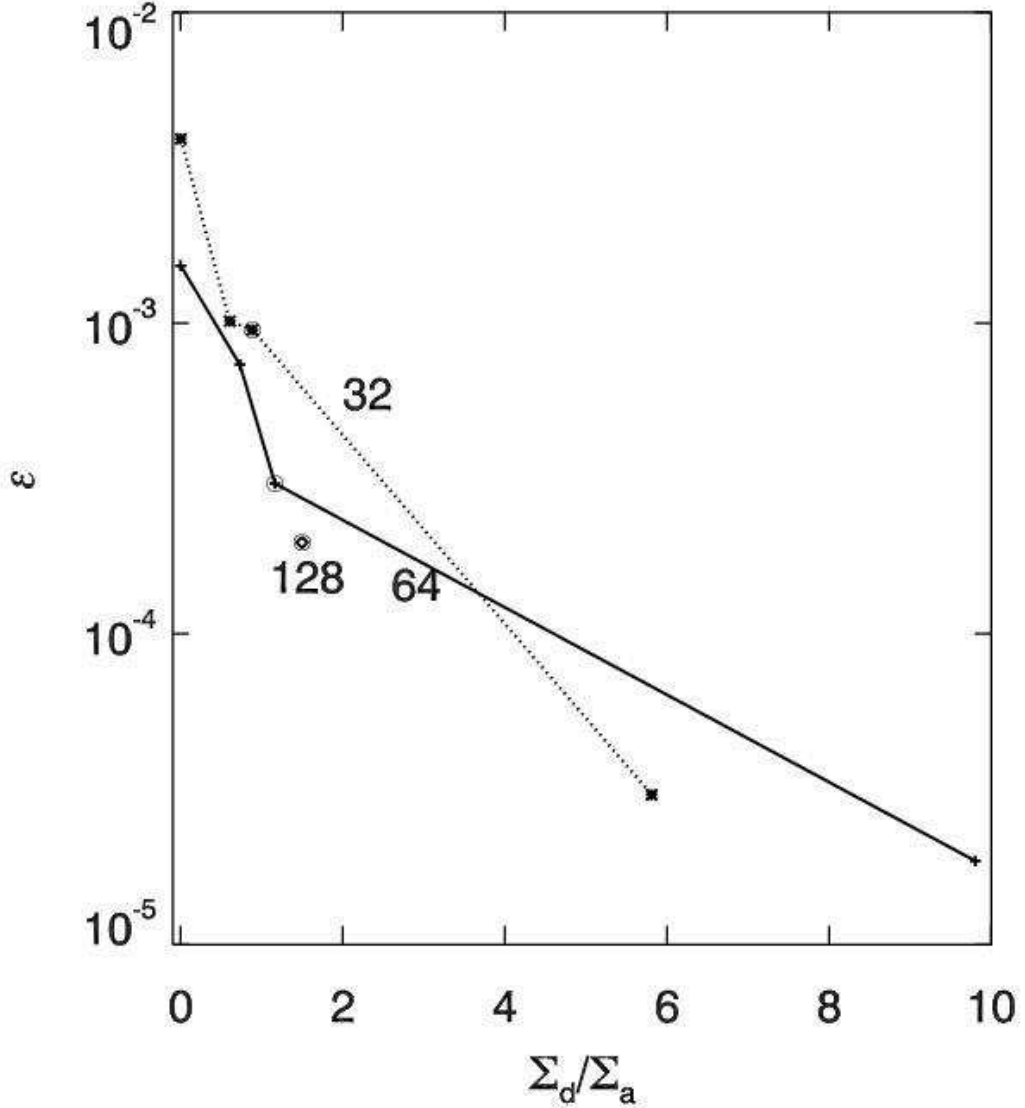}  
\caption{
\label{d_vs_deadsize}
Derived diffusion parameter $\epsilon$ estimated from turbulent
torques versus dead zone to active zone column density ratio
$\Sigma_D/\Sigma_A$ in all runs. Each resolution is labeled by its
$N_x = 32,64,128$, corresponding to our low, medium, and high
resolution runs. For the low and medium resolutions, there are several
$Re_M$ values; for the $N_x = 128$ case there is only one ($Re_M =
30$), marked by a diamond. The circled points correspond to the $Re_M
= 30$ case at each resolution, allowing a resolution study for $D$
versus $\Sigma_D/\Sigma_A$. The dead zone size should be more
resolution dependent towards the right of the figure, as the active
layers get thinner and thus increasing numbers of unstable modes are
inadequately resolved.}
\end{figure}

JGM06 collect all the uncertain properties of turbulence that need to
be derived from a numerical model into a single parameter
\begin{equation} \label{epsilon}
\epsilon = \frac{\tau_c}{\torb}
  \left(\frac{C_D}{0.046}\right)^2 = \frac{D}{(0.046)^2 r_p M_p},
\end{equation}
where the dimensionless coefficient 
\begin{equation}C_D = \frac{\langle \delta \Gamma^2
\rangle_t^{1/2}}{ 2\pi G \Sigma r_p M_p}.\end{equation} We calculate
$C_D$ using our simulated values of $F_y$, assuming fixed $r_p$, and
allowing $r_p$ and $M_p$ to cancel between $\delta \Gamma \propto r_p
M_p F_y$ and the denominator. Figure~\ref{d_vs_deadsize} shows
$\epsilon$ as a function of dead zone size in our various
simulations. We assume that the ratio of surface densities
$\Sigma_D/\Sigma_A$ mainly determines the strength of fluctuations in
$F_y$. This facilitates a comparison with the sizes of dead zones
predicted by ionization models \citep[e.g.][]{SMUN00,FTB02,IN06}.

Figure~\ref{d_vs_deadsize} includes a resolution study.  The circled
values show the same model at all three different resolutions, while
the two sets of models with extreme values of $Re_M$ were run at the
lower two resolutions. (Remember that the same value of $Re_M$
produces different dead zone thicknesses at different resolutions, as
shown in Figure~\ref{res_study}\emph{(a)}.) All of these comparisons
show that our derived diffusion parameter $\epsilon$ drops steadily with
increasing resolution, suggesting it is less well converged than the
average torque or dead zone thickness
(Fig.~\ref{res_study}\emph{(a)}). Thus, our estimates of the diffusion
coefficient and related quantities appear likely to be firm upper
limits to the true values.

JGM06 derive a fiducial value of $\epsilon \sim 0.5$ from the ideal
MHD simulations of \citet{N05}. Our local ideal MHD simulation, on the
other hand, gives a value of $\epsilon \simeq 10^{-3}$ at medium
resolution. On the other hand, JGM06 derive a fiducial correlation
time of $\sim 0.5$ from Nelson's simulations, which is quite close to
our $\sim 0.3$ value. This suggests that it is differences in the
torque magnitudes, rather than correlation times, that drives the
difference. The torque magnitudes are driven directly by the MRI
perturbations, which in turn may differ from N05 due to his lack of
stratification, by the difference in MRI strengths (his models have
$\alpha \simeq 5\times10^{-3}$, larger than ours by a factor of a
few), or by a difference in density perturbations between global and
local models. In our models with dead zones, we find $10^{-5} \lesssim
\epsilon \lesssim 10^{-3}$ over $3 \leq Re_{M} \leq \infty$.

In the disk model that we employ, even our ideal MHD model gives a
value of $\epsilon$ that suggests a terrestrial-mass planet in even
our most turbulent model will be in the advection dominated regime, as
shown in Figure~7 of JGM06. (Note that the flat parts of the curve in
that Figure correspond to advection.) However, even this reduced value
of $\epsilon$ is sufficient to drive diffusive migration for
protoplanets in the sub-Earth mass range.  For such planets, dead
zones may shift migration from diffusion to advection as they pass
radially in from a fully active region with stronger turbulence.

\subsection{Dead Zone Sizes}
Given the decrease of $\epsilon(J)$ with increasing dead zone size
that we find at all resolutions, it is reasonable to ask how close our
models come to a realistic dead zone thickness. Previous studies of
the ionization fraction in disks robustly show that the presence of
small dust grains ($< 1\, \mu \mathrm{m}$) with interstellar abundance
will cause dead zones to be orders of magnitude larger in column
density than their active zones \citep{SMUN00,SWH04,IN06}. Our $Re_M =
3$ models only reach a surface density ratio $\Sigma_D/\Sigma_A \simeq
10$, making dynamical models with significant submicron dust grains
well outside of current computational limits. However, once these dust
grains either sediment to the disk midplane or coagulate, ionization
occurs much more rapidly and the dead zone size shrinks. This may
occur in a small fraction of the disk lifetime, although a population
of new small grains with different properties could then be produced
by collisional processes \citep{DD05}.
For an $\alpha$-disk with $\dot{M} = 10^{-7} \mbox{ M}_{\odot}/\mbox{ yr}$ and
$\alpha = 10^{-2}$, the surface density ratios become
$\Sigma_D/\Sigma_A = 9.1$ at $1 \mbox{ AU}$ and 1.1 at~5~AU
\citep{IN06}. While this model assumes an $\alpha$-value somewhat
greater than we find, it suggests that simulations such as ours may
not be so far from realistic dead zone sizes. Our results indicate
that this ratio influences the magnitude of turbulent torques and thus
could determine the transition between advection- and
diffusion-dominated migration regimes for a variety of protoplanetary
masses.

\section{Discussion} \label{sec:discussion}

Our results indicate that in the thick dead zone that is expected
around 1--5~AU, turbulent torques will be reduced by at least two
orders of magnitude from those expected in a fully turbulent disk at
the same radii, allowing a uniform inward Type I migration mode to
become dominant even for protoplanets with $M_p \ll
\mathrm{M}_{\oplus}$.  A potential consequence of this is that if such
protoplanets form further out in the disk where the dead zone is
thinner, the inner, thick dead zone region of the disk could act as a
sink for protoplanets following random walks driven by the more
vigorous turbulence expected further out.

Because our simulations are local, we can only provide $\delta
\Gamma(t,J)$ at one value of angular momentum $J$ (that is, one radial
location in the disk). A radial gradient of the diffusion coefficient
$D(J)$ contributes to \emph{advection} and does not affect our main
results on dead zone diffusive migration. Indeed, a strong gradient in
the diffusion coefficient might be present at the radial boundary
between an active and dead zone in the midplane. Most models of
protoplanetary disk ionization suggest that there will be two such
boundaries, one in the outer disk where the column density increases
to the point that protostellar X-rays can no longer penetrate to the
midplane, and one in the inner disk where thermal ionization begins to
become important. JGM06 note that the orbital flux $F_J$ of the planet
distribution function $f$ can be written as
\begin{equation}
  F_J = (\overline{\Gamma} - \partial_J D) f - D \partial_J f,
\end{equation}
emphasizing the contribution of $\partial_J D$ to the effective mean
migration. In a region where protoplanets cross from an active region
to a dead zone in the midplane, $\epsilon$, and thus $D$
(eq.~\ref{epsilon}) changes by several orders of magnitude
(Fig.~\ref{d_vs_deadsize}) over a region of uncertain but likely
sub-AU width. At the inner boundary of the dead zone, $\partial_J D <
0$ contributing to outward mean migration, while at the outer
boundary, $\partial_J D > 0$, contributing to inward mean
migration. Thus, the random walk migration induced by a fully
turbulent disk tends to make the dead zone act as a trap for
sub-terrestrial mass protoplanets.

The magnitude of this effect is difficult to estimate from our
simulations because we ignore background gradients in the disk, and
thus our measurements of dead zone and fully active turbulent torques
are at equivalent radial positions in the disk. Nonetheless, future
global and statistical models should be able to predict if the effect
overcomes Type I migration in these regions strongly enough for the
dead zone to function as an efficient trap.

Recently, \citet*{TSD06} have performed simulations including a
dynamic $\eta(x,y,z,t)$ in which the dead zone disappears, at least
from the point of view of angular momentum transport: their results
suggest that while the MRI remains suppressed below $|z| \sim 0.5
\mathrm{H}$, there is still significant Maxwell stress from mean
fields that build up due to the vertical expulsion of flux and the
regeneration of azimuthal field from shear. They suggest that these
fields may grow large enough to invoke the \citet{T03} migration
mechanism in which magnetic resonances close to the planet may reverse
Type I migration. However, these results---while provocative---should
not significantly alter our main results. In our case, we isolate the
turbulent contribution to migration, ignoring the gravitational
back-reaction of a protoplanet on the disk. Any effects of the midplane
mean field, such as the \citet{T03} mechanism, are therefore absent by
construction. Indeed, since the MRI is suppressed at the midplane in
the \citet{TSD06} simulations, the turbulent overdensities will be
reduced, as in our simulations.

Our work depends on two major approximations: the restriction of our
simulation to the local, shearing box, and the adaptation of zero-mass
test particles fixed at the center of the box. Because of the latter
assumption, the possibility exists that our results underestimate
correlations because a moving particle would accelerate in the
direction of the largest current density fluctuation, thus increasing
the torque. However, the density structures will still have lifetimes
on the order of an orbit and they will not have different spatial
distributions. This means that in the lifetime of the fluctuation, the
protoplanet will not have time to move significantly from its current
position. Estimating the distance by an impulse approximation, the
distance moved over the lifetime of a perturbation is roughly $x/H
\sim 10^{-6}$--a trivial amount. Any global dynamical structures
associated with the MRI are absent from our models, including large
scale eddies that could change our results if they are sufficiently
massive. Large scale dynamical features may also affect the
correlation times of torques. Despite these caveats, we generally
expect our results to be qualitatively valid in demonstrating that
dead zones are regions of significantly reduced diffusive
migration. Dead zones may also have other consequences for planetary
migration such as those considered by \citet{MP03b} and \citet{MPT07}.

\section{Conclusions} \label{sec:conclusions}
We find that a dead zone in a protoplanetary disk can reduce the
magnitude of turbulent torque fluctuations on migrating protoplanets
by at least two orders of magnitude from a comparable fully ionized
region.  The thick dead zone expected at $1 \mbox{ AU}$, assuming a
dust-depleted disk, can have significant consequences for migration of
planets with $M_p \ll \mathrm{M}_{\oplus}$ driven by turbulent density
fluctuations.

The central assumptions of the stochastic diffusion model of
protoplanetary migration of JGM06 amenable to test in local
simulations seem well-founded. The torques have a finite correlation
time and have stationary distributions. These fluctuations maintain
nearly constant correlation times, independent of their amplitude.
The stochastic diffusion model suggests that if protoplanets form
outside of the dead zone and migrate in a diffusive manner, the dead
zone may act as a sink.

\acknowledgments JSO would like to thank the Max-Planck-Institut f\"ur
Astronomie for its kind hospitality during a portion of this work,
during which he was supported by an Annette Kade Graduate Student
Fellowship. M.K.R. Joung and R. Nelson made numerous helpful
suggestions for which we thank them. This work was partly supported by
the NASA Origins of Solar Systems Program under grant number
NNX07AI74G, as well as by the NSF under grant number
AST-0307793. Computations were performed on the Beehive Cluster at
Columbia University, the Parallel Computing Facility at AMNH, the Cray
XT3 ``Big Ben'' at the Pittsburgh Supercomputing Center, and the IBM
SP ``Datastar'' at the San Diego Supercomputing Center. The last two
are supported by the National Science Foundation.

\begin{deluxetable}{llllllllll}
\tablecaption{
\label{run_params}
Basic parameters for all runs.
}
\tablehead{
\colhead{Name} & \colhead{Resolution} & \colhead{$Re_M$\tablenotemark{a}} &
\colhead{$t_{max}/\torb$\tablenotemark{b}} & \colhead{$\Sigma_{A}/\Sigma_{D}$\tablenotemark{c}} &
\colhead{$\langle F_y^2\rangle_t^{0.5}$\tablenotemark{d}} &
\colhead{$\tau_c/\torb$\tablenotemark{e}} 
}
\startdata
32Rinf & $32\times 128^2$  & $\infty$ & 100 & -      & $5.17 (-3)$ & 0.31 \\
32R100 & $32\times 128^2$  & 100      & 100 & 1.65   & $2.63 (-3)$ & 0.31 \\
32R30  & $32\times 128^2$  & 30       & 200 & 1.13   & $2.27 (-3)$ & 0.39 \\
32R3   & $32\times 128^2$  & 3        & 100 & 0.177  & $5.17 (-4)$ & 0.24 \\
64Rinf & $64\times 256^2$  & $\infty$ & 100 & -      & $3.23 (-3)$ & 0.31 \\
64R100 & $64\times 256^2$  & 100      & 100 & 1.37   & $1.90 (-3)$ & 0.43 \\
64R30  & $64\times 256^2$  & 30       & 100 & 0.860  & $1.49 (-3)$ & 0.29 \\
64R3   & $64\times 256^2$  & 3        & 100 & 0.102  & $3.45 (-4)$ & 0.33 \\
128R30 & $128\times 512^2$ & 30       & 50  & 0.660  & $1.29 (-3)$ & 0.25 \\
\enddata
\tablenotetext{a}{Magnetic Reynolds number}
\tablenotetext{b}{Simulation time in orbits}
\tablenotetext{c}{Surface density ratio of active to dead zones}
\tablenotetext{d}{RMS y-component of force on protoplanet over orbits 25-100}
\tablenotetext{e}{Torque correlation time in orbits}
\end{deluxetable}

\begin{deluxetable}{llllllll}
\tablecaption{
\label{skew}
Skew for 10 orbit intervals from $25-95 \torb$ for $64 \times 256^2$
runs.  Columns are labeled by
the interval center time in orbits }
\tablehead{
\colhead{$Re_M$} & \colhead{30} & \colhead{40} & \colhead{50} & \colhead{60} & \colhead{70} & \colhead{80} & \colhead{90}
}
\startdata
$\infty$ & -0.283 &  0.463 &  0.076 & -0.157 &  0.209 & -0.026 &  0.028\\
100 &  0.175 & -0.158 & -0.505 & -0.113 & -0.170 & -0.651 & -0.163\\
30 & -0.240 &  0.305 &  0.277 & -0.123 &  0.254 & -0.680 &  0.455\\
3 &  0.097 &  0.278 &  0.673 &  0.145 &  0.048 & -0.001 &  0.076\\
\enddata
  
\end{deluxetable}

\begin{deluxetable}{llllllll}
\tablecaption{
\label{kurtosis}
Kurtosis for 10 orbit intervals from $25-95 \torb$ for $64 \times 256^2$
runs. Columns are labeled by
the interval center time in orbits. }
\tablehead{
\colhead{$Re_M$} & \colhead{30} & \colhead{40} & \colhead{50} & \colhead{60} & \colhead{70} & \colhead{80} & \colhead{90}
}
\startdata
$\infty$ &  0.657 &  1.446 &  0.460 &  0.379 &  0.555 &  1.025 &  1.274\\
100 &  0.910 &  0.304 &  0.425 &  0.064 &  0.675 &  0.944 &  0.439\\
30 &  2.066 &  1.133 &  1.057 &  0.201 &  0.573 &  3.112 &  1.223\\
3 &  0.362 &  0.746 &  2.883 &  0.923 &  0.921 &  1.386 &  0.258\\
\enddata
  
\end{deluxetable}

\begin{deluxetable}{llll}
\tablecaption{
\label{sano_density_relation}
Ratios of RMS density fluctuations and magnetic pressure.
}
\tablehead{\colhead{Run Name} & \colhead{$\langle \delta
  \rho^2 \rangle^{0.5}$ \tablenotemark{a}} &
\colhead{$\langle \delta P_{mag}^2\rangle^{0.5}/c_s^2$
  \tablenotemark{b}} & \colhead{$c_s^2 \langle\delta \rho^2\rangle^{0.5}/\langle\delta P_{mag}^2\rangle^{0.5}$ \tablenotemark{c}}
}
\startdata
32Rinf & $3.80 (-2)$ & $1.16 (-2)$ & 3.27 \\ 
32R100 & $2.51 (-2)$ & $5.93 (-3)$ & 4.24 \\
32R30  & $2.80 (-2)$ & $2.80 (-3)$ & 9.98 \\ 
32R3   & $6.84 (-3)$ & $3.12 (-4)$ & 21.9 \\ 
64Rinf & $4.27 (-2)$ & $1.93 (-2)$ & 2.22 \\ 
64R100 & $2.32 (-2)$ & $4.01 (-3)$ & 5.78 \\ 
64R30  & $1.74 (-2)$ & $2.28 (-3)$ & 7.63 \\ 
64R3   & $7.58 (-3)$ & $2.22 (-4)$ & 34.2 \\ 
128R30 & $1.64 (-2)$ &  $1.26 (-3)$ & 13.0\\
\enddata
\tablenotetext{a}{RMS density fluctuations at $t=50 t_{orb}$}
\tablenotetext{b}{variance of magnetic pressure normalized by $c_s^2$ at $t=50 t_{orb}$}
\tablenotetext{c}{ratio of cols a and b}

\end{deluxetable}

\clearpage

\end{document}